\documentclass[11pt]{article}
\usepackage{geometry}

\usepackage{hyperref}

\usepackage{setspace}
\usepackage{lscape}
\usepackage{graphicx}
\usepackage{epstopdf}
\usepackage{amsmath}
\usepackage{amsfonts}
\usepackage{float}
\usepackage{placeins}
\usepackage[cal=boondoxo]{mathalfa}
\usepackage{epstopdf}
\usepackage{epsfig}
\usepackage{multirow}
\usepackage{epstopdf}
\usepackage{rotating}
\usepackage{setspace}
\usepackage{parskip}
\usepackage[UKenglish]{babel}
\usepackage[UKenglish]{isodate}

\newcounter{bean}

\usepackage{authblk}
\begin{document}

\title{Fast Two-Stage Variational Bayesian Approach to Estimating Panel Spatial Autoregressive Models with Unrestricted Spatial Weights Matrices}

\author[1]{Deborah Gefang}
\author[2]{Stephen G. Hall}
\author[3]{George S. Tavlas}
\affil[1]{University of Leicester}
\affil[2]{University of Leicester, Bank of Greece and University of Pretoria}
\affil[3]{Bank of Greece and Hoover Institution, Stanford University }

\date{}

\maketitle

    \begin{abstract}
\noindent This paper proposes a fast two-stage variational Bayesian (VB) algorithm to estimate unrestricted panel spatial autoregressive models.
Using Dirichlet-Laplace priors, we are able to uncover
the spatial relationships between cross-sectional units without imposing any a priori
restrictions. Monte Carlo experiments show that our approach works well for
both long and short panels. We are also the first in the literature to develop
VB methods to estimate large covariance matrices with unrestricted sparsity
patterns, which are useful for popular large data models such as Bayesian
vector autoregressions. In empirical applications, we examine the spatial interdependence between euro area sovereign bond ratings and spreads. We find marked differences between
the spillover behaviours of the northern euro area countries and those of the
south.
%
%

    \end{abstract}

\doublespacing
\newpage

\section{Introduction}
The spatial autoregressive (SAR) models, first proposed by Cliff and Ord (1973), have been widely used in the literature to investigate the spatial dependence in cross-sectional
units (e.g., Anselin, 1988; Baltagi et al., 2003, 2013, Lee and Yu, 2010). In practice, the spatial weights matrices are usually set a priori based on geographical distances or
economic characters (e.g. Cliff and Ord, 1973, Anselin, 1988, Case, 1991). This is not surprising as a spatial weights matrix potentially involves $N^{2}-N$ interrelationships
between $N$ spatial units, which makes it difficult to estimate, especially when $N$ is large.

In recent years, a number of variable selection and parameter shrinkage methods have been developed to estimate the spatial weights matrices of panel SAR models. Among them,
methods resorting to least absolute shrinkage and selection operator (Lasso) of Tibshirani’s (1996) and its variants have gained a lot of attention. For example, Basak et al.
(2018) propose to estimate a triangular weights matrix under the assumption of recursive ordering. Ahrens and Bhattacharjee (2015) develop a two step Lasso estimator to identify
the weights matrix. Lam and Souza (2019) estimate the weights matrix using adaptive Lasso with sparse adjustment in mind. Most of the studies, however, usually impose sometime
unrealistic restrictions on the model's coefficients or covariances. Krock et al. (2021) develop a graphical Lasso approach to estimating the unrestricted covariances. Their
method, however, does not deal with the impacts of any possible exogenous variables. Moreover, to our knowledge, few of those studies focus on the short panels where $N$ is large
while $T$ is small. Only recently, Krisztin and Piribauer (2023) and Piribauer et al. (2023) use a hierarchical prior set-up to identify sparsity when $N$ exceeds $T$ by a large
margin, assuming the same spatial parameter for all the cross-sectional units.

This paper contributes to the SAR literature by developing a fast two-stage variational Bayesian (VB) approach to estimating panel SAR models with unknown spatial weights matrices.
We do not impose any restrictions on spatial weights matrix or the covariance functions, hence our approach lets the data speak. The prior we used for Bayesian regularization is
the Dirichlet–Laplace (D-L) prior of Bhattacharya et al. (2015). With D-L prior,  the entire posterior distribution concentrates at the optimal rate.  This nice feature remains
unchanged when the number of parameters to be estimated is much larger than the number of observations, providing strong theoretical justifications for the two-stage VB's
effectiveness in uncovering the spatial dependencies in a short panel.

Our second contribution is to develop VB methods to estimate large covariance matrices with a global-local shrinkage prior. We are among the first in the literature to develop VB
estimator for large covariance matrices with unknown sparsity patterns. Our VB methods using D-L prior can be easily extended to allow for other popular priors such as the graphic
Lasso of Wang (2012), the half-Cauchy prior of Makalic and Schimidt (2016)  and the graphical horseshoe prior of Li et al. (2019). This is not trivial as VB is a more
computationally efficient alternative of Markov Chain Monte Carlo (MCMC), and our approach can be used in estimating other popular models involving large covariance matrices such
as large Bayesian vector autoregressions (BVARs).\footnote{Matlab code is provided.}

We have conducted a wide range of simulation studies using a traditional panel SAR model and a panel SAR model that takes account of the simultaneous relationships between
cross-sectional groups.\footnote{This research used the ALICE High Performance Computing Facility at the University of Leicester.} Monte Carlo experiments show that two-stage VB is accurate and computationally efficient when $T\gg N$, which usually is more pertinent to macroeconomic and
financial data. When $N\gg T$, which tends to be more relevant to microeconomic data, two-stage VB estimates tend to have slightly larger biases and empirical standard deviations.
Tighter priors can help lessen that problem.

In empirical applications, we use two-stage VB to estimate a two-equation simultaneous spatial model using panel data comprising sovereign bond ratings and spreads of ten eurozone
countries. The research is motivated by the importance of a more in-depth understanding in how cross-country contagion works in the southern euro area countries and their northern
counterparts, especially in the European debt crises (e.g. Gibson et al., 2021, Hall et al., 2022). Using rolling windows of 2 years, we compare and contrast how a rate or spread
change in the south and the north affect each individual member country. Our results provide ample evidence for the marked differences between the two country groups, both in how a
shock to the south and a shock to the north impact a particular country in a very different way and in how the shocks to a particular country group affect a southern and a northern
country very differently. Our results also highlight the big impacts of the global financial crises and how the south was severely affected by the European debt crises while the
north was much less so.

The rest of the paper is organised as follows. Section 2 extends the traditional panel SAR models to an unrestricted panel SAR. Section 3 develops the two-stage VB. Section 4
conducts Monte Carlo studies. Section 5 applies two-stage VB to the sovereign bond ratings and spreads data of ten euro area countries. Section 6 concludes. Sources for data used
in the empirical example are listed in the Appendix. Detailed VB derivation formulas and more extensive Monte Carlo results are relegated to Online Supplements.\footnote{Online Supplements can be found at https://github.com/DBayesian/GKT2022}

\section{Unrestricted Panel SAR Model}

In this section, we start from a traditional standard panel SAR and then relax the restrictions imposed upon it in steps, with the aim of giving a flavour of the differences
between the traditional model and the unrestricted panel SAR model that we set to estimate using two-stage VB.

Let $Y$, $X$ and $V$ denote the $T\times N$ matrix of endogenous variables,  $T\times (Nm)$ matrix of exogenous variables, and $T\times N$ matrix of disturbances, respectively. A
traditional panel SAR model takes the following form:

\begin{equation}\label{sar_indv}
 y_{t}=\lambda W_{n}y_{t}+X_{t}\beta+u_{t},  \hspace{0.2cm} |\lambda|<1
\end{equation}
where $y_{t}=(Y_{t1},Y_{t2},...,Y_{tN})'$ is the $N\times 1$ vector of observations of the dependent variables, $W_{n}$ is an $N\times N$ known spatial weights matrix with zero
diagonal entries, $X_{t}=\left(
                                \begin{array}{c}
                                  X_{t,1} \\
                                X_{t,2} \\
                                  ... \\
                                   X_{t,N}\\
                                \end{array}
                              \right)
$ is the $N\times m$ matrix of exogenous variables, with $X_{t,i}$ denoting the $1\times m$ row vector of exogenous variables associated with dependent variable $y_{ti}$, $\beta$
is an $m\times 1$ vector of parameters, $\lambda$ is a scalar parameter, and $u_{t}=(V_{t1},V_{t2},...,V_{tN})'$ is the $N\times 1$ i.i.d error terms with mean zero and diagonal
covariance matrix $ \left(
                                                      \begin{array}{cccc}
                                                        \sigma^{2} & 0 & ... & 0 \\
                                                        0& \sigma^{2} & ... & 0 \\
                                                        .. & .. & ... & .. \\
                                                       0 &0 & ... & \sigma^{2} \\
                                                      \end{array}
                                                    \right)
$.

Model (\ref{sar_indv}) imposes the following unrealistic restrictions on the data generating process: 1) $W_{N}$  is predetermined,  in a fashion that is not related to the
variations in the data; 2) $\lambda$ and $\beta$ remain the same across equations associated with different dependent variables; and 3) the covariance matrix  of  $u_{t}$ is
diagonal with the same diagonal entries.

 Relaxing those restrictions, model (\ref{sar_indv}) can be written as:

\begin{equation}\label{sar_indv2}
 y_{t}=\tilde{\lambda} \odot (\widetilde{W}_{N}y_{t})+\widetilde{X}_{t}\tilde{\beta}+u_{t},  \hspace{0.2cm} |\lambda_{b}|_{\infty}<1
\end{equation}
where $\widetilde{W}_{N}$ is an $N\times N$ unknown spatial weight matrix with zero diagonal entries, $\tilde{\lambda}$ is a $N\times 1$ parameter vector,  $\odot$ is the Hadamard
product, $\widetilde{X}_{t}=\left(
                     \begin{array}{cccc}
                       X_{t,1} & 0 & ... & 0 \\
                      0 & X_{t,2} & ... & 0 \\
                       .. & .. & ... & .. \\
                       0 & 0 & ... & X_{t,N} \\
                     \end{array}
                   \right)
$, $\tilde{\beta}$ is a $Nm\times 1$ parameter vector, and $u_{t}$ is i.i.d with mean zero and diagonal covariance matrix $\tilde{\Sigma}$ with diagonal entries that can be
different from each other. In this model, the dimensions of $X_{t,i}$ and $X_{t,j}$  for $i\neq j$ can differ from each other. Let the dimension of $X_{t,i}$ to be $1\times m_{i}$.
The dimension of the parameter vector $\tilde{\beta}$ is thus $(\sum_{i=1}^{n}m_{i})\times 1$.

Note that model (\ref{sar_indv2}) is quite flexible. For example, with appropriate restrictions, it can be easily transformed back into the traditional form described in
(\ref{sar_indv}) or a panel SAR containing the simultaneous cross-sectional spatial relationship  as described in Yang and Lee (2017) and  Liu and Saraiva (2019).

Since our main concerns in panel SAR models are the spillover effects, there is therefore little research interest in separately identifying $\tilde{\lambda}$ and
$\widetilde{W}_{N}$. What we care about is the product $(\tilde{\lambda} \otimes l_{N}) \odot \widetilde{W}_{N}$, where $l_{N}$ is a $N\times 1$ column of ones and $\otimes$ is the
Kronecker product, as $(\tilde{\lambda} \otimes l_{N}) \odot \widetilde{W}_{N}$ is the $N\times N$ parameter matrix which captures the spillover effects between spatial units.

Let $\Lambda= (\tilde{\lambda} \otimes l_{N}) \odot \widetilde{W}_{N}$.   Model (\ref{sar_indv2}) can be written as:
\begin{equation}\label{sem_indv}
 y_{t}= \Lambda y_{t}+\widetilde{X}_{t}\tilde{\beta}+u_{t},
\end{equation}
where $I_{N}-\Lambda$ is nonsingular and the characteristic roots of $I_{N}-\Lambda$ lie within the unit circle.

The attractiveness of model (\ref{sem_indv}) is that it turns an unrestricted panel SAR model into a system of simultaneous equations (SEM). As shown in Zellner and Thell (1962)
and  Fox (1979), the $i^{th}$ equation in model (\ref{sem_indv}) is just identified if $ \sum_{i=1}^{N}m_{i}=N-1+m_{i}$ and over-identified if $\sum_{i=1}^{N}m_{i}>N-1+m_{i}$.
Under these circumstances, a myriad of estimation methods, such as two-stage least squares (2SLS), three-stage least squares (3SLS), maximum likelihoods methods and simultaneous
generalized method (GMM), can be used to uncover the structural parameters $\Lambda$ and $\tilde{\beta}$. Moreover, standard tests can be developed to test the restrictions on
$\tilde{\lambda}$, $\widetilde{W}_{N}$ and $\tilde{\beta}$, if those restrictions are of the researchers' concerns.

 This paper proposes to estimate $ \Lambda $ and $\tilde{\beta}$ in two stages as it is computationally simple. To estimate the parameters associated with the $i^{th}$ individual
 dependent variable, in the first stage, we estimate
 \begin{equation}\label{1st_stage}
 Y_{/i}= X\Upsilon_{i}+E_{i},
\end{equation}
where $ Y_{/ i}$ is the $T\times (N-1)$ matrix of dependent variables except for the $i^{th}$ dependent variable, and $E_{i}$ is a $T\times (N-1)$ matrix of error terms whose
precision matrix might not be diagonal.

Making use of the estimated $\Upsilon_{i}$, in the second stage, we estimate
\begin{equation}\label{2st_stage}
 y_{i}=\widehat{Y}_{ /i}(\Lambda _{i\bullet})'+X_{\bullet, i}\tilde{\beta}_{i}+u_{i},
\end{equation}
where $ y_{i}$ is the $T\times 1$ vector of the $i^{th}$ dependent variable,  $\widehat{Y}_{/i}=X\Upsilon_{i}$, $\Lambda _{i\bullet}$ is the $1\times (n-1) $ vector of the $i^{th}$
row of $\Lambda$ with $\Lambda_{ii}$ dropped, and $\tilde{\beta}_{i}$ is the corresponding coefficients in $\tilde{\beta}$.

Note that in a panel SAR model, we can have $N\gg T$ and $(N-1+m_{i})\gg T$, which makes it difficult or even impossible to uncover $B_{i}$ and $\Lambda _{i\bullet}$ using
traditional estimation techniques.

\section{Two-stage VB}
As detailed in Ormerod and Wand (2010) and  Blei et al. (2017),  the essence of VB is to use appropriate densities from a mean field variational family to approximate the posterior
densities through minimizing the Kullback-Leibler divergence, which is equivalent to maximising the evidence lower bound (ELBO). As a more efficient alternative to MCMC,  VB has
been increasingly used in sophisticated models involving large data where MCMC is too computationally expensive or even untenable (e.g. Gefang et al.,  2020, 2022, Loaiza-Maya et
al., 2022).

In the two-stage VB, we identify the parameters in model (\ref{sem_indv}) equation by equation. In each stage, we update the parameters using the approximate $q$ densities by
iterations. The convergence of the algorithm can be measured by the changes in ELBO  across iterations is less than a convergence criteria. When the number of parameters is large,
however, calculating ELBO can be time consuming. It is therefore more convenient to check if convergence has occurred by examining if the VB estimates of parameters stop changing
across iterations.


\subsection{First-stage VB}

In the first stage, we estimate model (\ref{1st_stage}) in order to construct the predicted value of $ Y_{/i}$.

Let $\gamma=vec(\Upsilon_{i})$. We set hierarchical D-L prior for the $j^{th}$, for $j=1,...,np$, element of $\gamma$ as follows:
\begin{equation}
\gamma_{j}|\phi,\tau\sim DE(\phi_{j}\tau), \hspace{0.2cm} \phi_{j}\sim Dir(a,...,a), \hspace{0.2cm} \tau\sim G(npa,1/2).
\label{DE_L}
\end{equation}
where $DE(\bullet)$ denotes Double Exponential or Lapalace distribution, $Dir(\bullet)$ denotes Dirichlet distribution, $G(\bullet)$ denotes Gamma distribution, $n=N-1$, and
$p=\sum_{i=1}^{N}m_{i}$.

Next, we set Exponential priors and D-L priors for the elements of $\Omega$, the precision matrix of $E_{i}$, as follows:
\begin{equation}\label{Omg_priors}
  \begin{aligned}
    &\omega_{ii} \sim Exp(\underline{s}), \hspace{0.2cm} i=1,...,n \\
    &\omega_{ij} \sim N(0,\psi_{\omega,ij}\phi_{\omega,i j}^{2}\tau_{\omega}^{2}), \hspace{0.2cm} \psi_{\omega,ij}\sim Exp(1/2),\hspace{0.2cm} i<j=2,....,n, \\
    & \phi_{\omega,ij}\sim Dir(a_{\omega},....,a_{\omega}), \hspace{0.2cm} \tau_{\omega}\sim G(\frac{n^{2}-n}{2}a_{\omega},1/2)
  \end{aligned}
\end{equation}
where, with a slight abuse of notations, we use $\omega_{ii}$ and $\omega_{ij}$ to denote the diagonal and off-diagonal elements of $\Omega$.

Following Wang's (2012) Block Gibbs sampler to update the relevant parameters and hyperparameters,  we use the last column and row of $\Omega$ as an example on how to update
$\Omega$.

Let $S=E_{i}'E_{i}$ and $H$ be the $n\times n$ matrix with $0$ diagonal elements and the off diagonal element at $i^{th}$ row and $j^{th}$ column be
$\psi_{\omega,ij}\phi_{\omega,ij}^{2}\tau_{\omega}^{2}$. Partition $\Omega$, $S$ and $H$ as follows:

\begin{equation}\label{part}
 \Omega =\left(
           \begin{array}{cc}
             \Omega_{-n,-n} &  \omega _{-n,n} \\
              \omega_{-n,n}^{'} & \omega_{nn} \\
           \end{array}
         \right)
 \hspace{0.2cm}
 S =\left(
           \begin{array}{cc}
             S_{-n,-n} &  s_{-n,n} \\
              s_{-n,n}^{'} & s_{nn} \\
           \end{array}
         \right)
 \hspace{0.2cm}
 H =\left(
           \begin{array}{cc}
             H_{-n,-n} &  h_{-n,n} \\
              h_{-n,n}^{'} & h_{nn} \\
           \end{array}
         \right)
\end{equation}
where $-n$ denotes the set of all indices except for $n$.

Relegating technical details to Online Appendix A, we outline the VB approximation densities as follows:

\subsubsection{$q(\gamma)$}
\begin{equation}\label{q_beta}
  q(\gamma)\sim N(\bar{\gamma}, \overline{V}),
\end{equation}
where
 $$\overline{V}=(V^{-1}+\overline{\Omega}\otimes(X'X))^{-1},$$
$$\overline{\gamma}=\overline{V}(\overline{\Omega}\otimes X')vec(Y_{/i}), $$
and
 $$V^{-1}=diag(\frac{1}{\overline{\psi_{1}}\hspace{0.1cm}\overline{\phi_{1}^{2}}\hspace{0.1cm}\overline{\tau^{2}}},...,\frac{1}{\overline{\psi_{np}}\hspace{0.1cm}\overline{\phi_{np}^{2}}\hspace{0.1cm}\overline{\tau^{2}}})$$

 \subsubsection{$q(\tau)$}

\begin{equation}\label{qtau}
  q(\tau)\sim giG[npa-np,1,\sum_{j=1}^{np}2(\overline{\gamma}_{j}^{2}+\overline{V}_{jj})^{1/2}\frac{1}{\overline{\phi_{j}}}],
\end{equation}
Let $\chi=\sum_{j=1}^{np}2(\overline{b}_{j}^{2}+\overline{V}_{jj})^{1/2}\frac{1}{\overline{\phi_{j}}}$, we have
$$\bar{\tau}=\frac{\sqrt{\chi}K_{npa-np+1}(\sqrt{\chi})}{K_{npa-np}(\sqrt{\chi})}$$
$$\overline{\tau^{2}}=\bar{\tau}^{2}+\chi[\frac{K_{npa-np+2}(\sqrt{\chi})}{K_{npa-np}(\sqrt{\chi})}-(\frac{K_{npa-np+1}(\sqrt{\chi})}{K_{npa-np}(\sqrt{\chi})})^{2}]$$
where $K_{\ast}[\bullet]$ is the modified Bessel functions of the second kind.

\subsubsection{$q(\psi_{j})$}

\begin{equation}\label{qpsi}
  q(\psi_{j}^{-1})\sim iG(\sqrt{\frac{\overline{\phi_{j}^{2}}\hspace{0.2em} \overline{\tau^{2}}}{\overline{\gamma}_{j}^{2}+\overline{V}_{jj}}},1),
\end{equation}
where $iG(\bullet)$ denotes s Inverse Gaussian distribution.

Let $\rho=\sqrt{\frac{\overline{\phi_{j}^{2}}\hspace{0.2em} \overline{\tau^{2}}}{\overline{\gamma}_{j}^{2}+\overline{V}_{jj}}}$,
$$\overline{\psi_{j}^{-1}}=\rho$$
and
$$\overline{\psi}_{j}=1+1/\rho$$

\subsubsection{$q(\phi_{j})$}

\begin{equation}\label{qxi}
q(\xi_{j})\sim giG(a-1,1,2 \sqrt{\overline{\gamma}_{j}^{2}+\overline{V}_{jj}} ),
\end{equation}
where $giG(\bullet)$ denotes the generalized inverse Gaussian distribution.

Let $\varpi=2 \sqrt{\overline{\gamma}_{j}^{2}+\overline{V}_{jj}}$, we have
$$
\overline{ \xi_{j}}=\frac{\sqrt{\varpi}K_{a}
(\sqrt{\varpi})}
{K_{a-1}(\sqrt{\varpi})},
$$
and
$$var(\xi_{j})=\varpi
\{\frac{K_{a+1}(\sqrt{\varpi})}
{K_{a-1}(\sqrt{\varpi})}
-[\frac{K_{a}(\sqrt{\varpi})}
{K_{a-1}(\sqrt{\varpi})}]^{2}\},$$
where $var(\bullet)$ denotes the variance.

Scaling $\xi_{i}$, we have
$$\overline{\phi_{j}}=\frac{\overline{ \xi_{j}}}{\sum_{j=1}^{np}\overline{ \xi_{j}}},$$
and
$$\overline{\phi_{j}^{2}}=\overline{\phi_{j}}^{2}+\frac{var(\xi_{j})}{(\sum_{j=1}^{np}\overline{ \xi_{j}})^{2}}$$

Thus, the optimal $q$ density of $\phi_{i,j}$ takes the following form:

\begin{equation}\label{qphii}
q( \phi_{j})\sim giG(a-1,\sum_{j=1}^{np}\overline{ \xi_{j}},\frac{2\sqrt{\overline{\gamma}_{j}^{2}+(\overline{V}_{jj})^{2}}}{\sum_{j=1}^{np}\overline{ \xi_{j}}})
\end{equation}

\subsubsection{$q(b_{1})$}
Let $b_{1}=\omega_{n,n}- \omega _{-n,n}' \Omega_{-n,-n}^{-1} \omega_{-n,n}$ .
\begin{equation}\label{qb1aaa}
q(b_{1})\sim G(\frac{T}{2}, \overline{s}_{n,n}),
\end{equation}

where
$$\overline{s}_{n,n}=\frac{1}{2}(s_{n,n}+tr(X'X\overline{V_{n}})+\underline{s}),$$
and
$$\overline{V_{n}}=V_{(n-1)\times p+1:n\times p, (n-1) \times p+1:n \times p}.$$

Hence
$$\overline{b_{1}}=\frac{\frac{T}{2}}{\overline{s}_{n,n}}$$

\subsubsection{$q(b_{2})$}
Here we use $b_{2}$ to denote $\omega _{-n,n}$. Let $\overline{s}_{-n,n}=s_{-n,n}+\tilde{s}_{-n,n}$, where $\tilde{s}_{-n,n}$ is a $(n-1)\times 1$ vector with the $j^{th}$ element
being $tr(X'XA_{j})$ and $A_{j}=V_{(j-1)\times p+1:j \times p,(j-1)\times p+1:j \times p }$.

\begin{equation}\label{qb2aaa}
q(b_{2})\sim N(-\overline{C}\overline{s}_{-n,n},\overline{C}),
\end{equation}
where
$$\overline{C}=(2\overline{s}_{n,n}\Omega_{-n.-n}^{-1}+\overline{H}^{*-1})^{-1},$$
and
$$\overline{b_{2}}=(-\overline{C}\overline{s}_{-n,n}).$$

Note that $\overline{H}^{*}=diag(\overline{h}_{-n,n})$, and  $j^{th}$ element of $\overline{h}_{-n,n}$ is
$\overline{\psi_{\omega,jn}}\overline{\phi_{\omega,jn}^{2}}\overline{\tau_{\omega}^{2}}$.

\subsubsection{$q(\tau_{\omega})$}

\begin{equation}\label{qtauOmg}
  q(\tau_{\omega})\sim giG[\frac{n^{2}-n}{2}(a_{\omega}-1),1,\sum_{j<k}2(\overline{\omega_{jk}}^{2}+\overline{C}_{jj})^{1/2}\frac{1}{\overline{\phi_{\omega,jk}}}],\footnote{Note
  that $\overline{C}_{jj}$ changes when $k$ changes.}
\end{equation}
Let $\chi_{\omega}=\sum\limits_{j<k}2(\overline{\omega_{jk}}^{2}+\overline{V}_{jj})^{1/2}(\overline{\phi_{\omega,jk}})^{-1}$, we have
$$\overline{\tau_{\omega}}=\frac{\sqrt{\chi_{\omega}}K_{\frac{n^{2}-n}{2}(a_{\omega}-1)+1}(\sqrt{\chi_{\omega}})}{K_{\frac{n^{2}-n}{2}(a_{\omega}-1)}(\sqrt{\chi_{\omega}})}$$

$$\overline{\tau_{\omega}^{2}}=\overline{\tau_{\omega}}^{2}+\chi_{\omega}[\frac{K_{\frac{n^{2}-n}{2}(a_{\omega}-1)+2}(\sqrt{\chi})}{K_{\frac{n^{2}-n}{2}(a_{\omega}-1)}(\sqrt{\chi_{\omega}})}
-(\frac{K_{\frac{n^{2}-n}{2}(a_{\omega}-1)+1}(\sqrt{\chi_{\omega}})}{K_{\frac{n^{2}-n}{2}(a_{\omega}-1)}(\sqrt{\chi_{\omega}})})^{2}]$$

\subsubsection{$q(\psi_{\omega,jn})$}
\begin{equation}\label{qpsiOmg}
  q(\psi_{\omega,jn}^{-1})\sim iG(\sqrt{\frac{\overline{\phi_{\omega,jn}^{2}}\hspace{0.2em} \overline{\tau_{\omega}^{2}}}{\overline{\omega_{jn}^{2}}+\overline{C}_{jj}}},1),
\end{equation}

Let $\rho_{\omega}=\sqrt{\frac{\overline{\phi_{\omega,jn}^{2}}\hspace{0.2em} \overline{\tau_{\omega}^{2}}}{\overline{\omega_{jn}}^{2}+\overline{C}_{jj}}}$,
$$\overline{\psi_{\omega,jn}^{-1}}=\rho_{\omega}$$
and
$$\overline{\psi_{w,jn}}=1+1/\rho_{\omega}$$

\subsubsection{$q( \phi_{\omega,jn})$}

\begin{equation}\label{qxiOmg}
q(\xi_{\omega,jn})\sim giG(a_{\omega}-1,1,2 \sqrt{\overline{\omega_{jn}}^{2}+\overline{C}_{jj}} )
\end{equation}

Let $\varpi_{\omega}=2 \sqrt{\overline{\omega_{jn}}^{2}+\overline{C}_{jj}}$, we have
$$
\overline{ \xi_{\omega,jn}}=\frac{\sqrt{\varpi_{\omega}}K_{a_{\omega}}
(\sqrt{\varpi_{\omega}})}
{K_{a_{\omega}-1}(\sqrt{\varpi_{\omega}})},
$$
and
$$var(\xi_{\omega,jn})=\varpi_{\omega}
\{\frac{K_{a_{\omega}+1}(\sqrt{\varpi_{\omega}})}
{K_{a_{\omega}-1}(\sqrt{\varpi_{\omega}})}
-[\frac{K_{a_{\omega}}(\sqrt{\varpi_{\omega}})}
{K_{a_{\omega}-1}(\sqrt{\varpi_{\omega}})}]^{2}\},$$
where $var(\bullet)$ denotes the variance.

Scaling $\xi_{\omega,jn}$, we have
$$\overline{\phi_{\omega,jn}}=\frac{\overline{ \xi_{\omega,jn}}}{\sum_{j<k}\overline{ \xi_{\omega,jk}}},$$

and
$$\overline{\phi_{\omega,jn}^{2}}=\overline{\phi_{\omega,jn}}^{2}+\frac{var(\xi_{\omega,jn})}{(\sum\limits_{j<k}\overline{ \xi_{\omega,jk}})^{2}}$$

Thus, the optimal $q$ density of $\phi_{\omega,ij}$ takes the following form:

\begin{equation}\label{qphiiOmg}
q( \phi_{\omega,jn})\sim giG(a-1,\sum\limits_{j<k}\overline{ \xi_{\omega,jk}},\frac{2\sqrt{\overline{\omega_{jn}}^{2}+(\overline{C}_{jj})^{2}}}{\sum\limits_{j<k}\overline{
\xi_{\omega,jk}}})
\end{equation}

\subsection{Second-stage VB}
We explain the technical details of second-stage VB in Online Appendix B. Below we briefly describe the priors of the parameters and hyperparameters and then  provide their optimal
$q$ densities.

Let $Z=(\widehat{Y}_{ /i} \hspace{0.2cm}X_{\bullet, i})'$ and $\theta=[(\Lambda _{i\bullet})' \hspace{0.2cm}\tilde{\beta}_{i}]$.

We elicit hierarchical D-L prior for $\theta$ as follows:
\begin{equation}
\theta_{j}|\tilde{\phi},\tilde{\tau}\sim DE(\tilde{\phi}_{j}\tilde{\tau}), \hspace{0.2cm} \tilde{\phi}_{j}\sim Dir(\tilde{a},...,\tilde{a}), \hspace{0.2cm}\tau\sim
G(k\tilde{a},1/2)
\label{DE_L_2st}
\end{equation}
where $k=N-1+m_{i}$.

Next we set a Gamma prior for $\sigma^{-2}$:
\begin{equation}
\sigma^{-2}\sim G(\nu,\tilde{S}).
\label{Sigma_2st}
\end{equation}

The VB optimal densities can be found as follows:

\subsubsection{$q(\theta)$}

\begin{equation}\label{qtheta}
  q(\theta)\sim N(\overline{\theta},\overline{\tilde{V}}),
\end{equation}
where $$\overline{\tilde{V}}=(\frac{\frac{T}{2}+\nu}{\overline{\tilde{S}}}Z'Z+\tilde{V}^{-1})^{-1}$$
$$\overline{\theta}=(\frac{\frac{T}{2}+\nu}{\overline{\tilde{S}}})\overline{\tilde{V}}Z'y_{i}$$
$$\tilde{V}^{-1}=diag(\overline{\tilde{\psi}_{1}^{-1}}\hspace{0.1cm} \overline{\tilde{\phi}_{1}^{-2}}\overline{\tilde{\tau}^{-2}},...,(\overline{\tilde{\psi}_{k}^{-1}}
\hspace{0.1cm} \overline{\tilde{\phi}_{k}^{-2}}\overline{\tilde{\tau}^{-2}})$$

\subsubsection{$q(\sigma^{-2})$}

\begin{equation}\label{qsigma}
  q(\sigma^{-2})\sim G(\frac{T}{2}+\nu,\overline{\tilde{S}}),
\end{equation}
where
$$\overline{S}=\frac{1}{2}[||y_{i}-Z\overline{\theta}||^{^{2}}+tr(Z'Z\overline{\tilde{V}})]+\tilde{S}$$

\subsubsection{$q(\tilde{\tau})$}

\begin{equation}\label{qtau}
  q(\tilde{\tau})\sim giG[k\tilde{a}-k,1,\sum_{j=1}^{k}2(\overline{\theta}_{j}^{2}+\overline{\tilde{V}}_{jj})^{1/2}\frac{1}{\overline{\tilde{\phi}_{j}}}],
\end{equation}
Let $\tilde{\chi}=\sum_{j=1}^{k}2(\overline{\theta}_{j}^{2}+\overline{\tilde{V}}_{jj})^{1/2}\frac{1}{\overline{\tilde{\phi}_{j}}}$, we have
$$\overline{\tilde{\tau}}=\frac{\sqrt{\tilde{\chi}}K_{k\tilde{a}-k+1}(\sqrt{\tilde{\chi}})}{K_{k\tilde{a}-k}(\sqrt{\tilde{\chi}})}$$
and
$$\overline{\tilde{\tau}^{2}}=\overline{\tilde{\tau}}^{2}+\chi[\frac{K_{k\tilde{a}-k+2}(\sqrt{\tilde{\chi}})}{K_{k\tilde{a}-k}(\sqrt{\tilde{\chi}})}-(\frac{K_{k\tilde{a}-k+1}(\sqrt{\tilde{\chi}})}{K_{k\tilde{a}-k}(\sqrt{\tilde{\chi}})})^{2}]$$


\subsubsection{$q(\tilde{\psi}_{j})$}

\begin{equation}\label{qpsi}
  q(\frac{1}{\tilde{\psi}_{j}})\sim iG(\sqrt{\frac{\overline{\tilde{\phi}_{j}^{2}}\hspace{0.2em}
  \overline{\tilde{\tau}^{2}}}{\overline{\theta}_{j}^{2}+\overline{\tilde{V}}^{jj}}},1),
\end{equation}

Let $\tilde{\rho}=\sqrt{\frac{\overline{\tilde{\phi}_{j}^{2}}\hspace{0.2em} \overline{\tilde{\tau}^{2}}}{\overline{\theta}_{j}^{2}+\overline{\tilde{V}}^{jj}}}$,
$$\overline{\frac{1}{\tilde{\psi}_{j}}}=\tilde{\rho}$$
and
$$\overline{\tilde{\psi}}_{j}=1+1/\tilde{\rho}$$

\subsubsection{$q( \tilde{\phi}_{j})$}

\begin{equation}\label{qxi}
q( \tilde{\xi}_{j})\sim giG(\tilde{a}-1,1,2\sqrt{\overline{\theta}_{j}^{2}+(\overline{\tilde{V}}_{jj})^{2}})
\end{equation}
Let $\tilde{\varpi}=2\sqrt{\overline{\theta}_{j}^{2}+(\overline{\tilde{V}}_{jj})^{2}}$, we have
$$
\overline{ \tilde{\xi}}_{j}=\frac{\sqrt{\tilde{\varpi}}K_{\tilde{a}}
(\sqrt{\tilde{\varpi}})}
{K_{\tilde{a}-1}(\sqrt{\tilde{\varpi}})},
$$
and
$$var(\tilde{\xi}_{j})=\tilde{\varpi}
\{\frac{K_{\tilde{a}+1}(\sqrt{\tilde{\varpi}})}
{K_{\tilde{a}-1}(\sqrt{\tilde{\varpi}})}
-[\frac{K_{\tilde{a}}(\sqrt{\tilde{\varpi}})}
{K_{\tilde{a}-1}(\sqrt{\tilde{\varpi}})}]^{2}\}.$$

Scaling $\tilde{\xi}$, we have
$$\overline{\tilde{\phi}}_{j}=\frac{\overline{ \tilde{\xi}}_{j}}{\sum^{k}\overline{ \tilde{\xi}}_{j}},$$
and
$$\overline{\tilde{\phi}_{j}^{2}}=\overline{\tilde{\phi}}_{j}^{2}+\frac{var(\tilde{\xi}_{j})}{(\sum^{k}\overline{ \tilde{\xi}}_{j})^{2}}$$

Thus, the optimal $q$ density of $\tilde{\phi}_{j}$ takes the following form:

\begin{equation}\label{qphii}
q( \tilde{\phi}_{j})\sim giG[\tilde{a}-1,\sum^{k}\overline{ \tilde{\xi}}_{j},(2\sqrt{\overline{\theta}_{j}^{2}+(\overline{\tilde{V}}_{jj})^{2}})/(\sum^{k}\overline{
\tilde{\xi}}_{j})]
\end{equation}

\section{Monte Carlo Studies}
In the Monte Carlo studies, we look into two traditional panel SAR models of various sample sizes.
The first model is:
\begin{equation}\label{simuSAr}
  y_{t}=0.6W_{N}y_{t}+0.9x_{t}+u_{t},
\end{equation}
where to specify $W_{N}$,  we let each cross-sectional unit be connected with the unit ahead of it and the unit behind, and then normalize $W$ by rows. When conducting Monte Carlo,
we generate $x_{t}$ and $u_{t}$ independently from $N(0,1)$ and $0.1N(0,1)$, respectively.

The second model is:
 \begin{equation}\label{liu_sa}
  \begin{aligned}
& y_{t,1} = 0.5y_{t,2}+0.6W_{N_{1}}y_{t,1}+0.4W_{N_{2}}y_{t,2}+0.9x_{t,1}+u_{t,1}\\
&y_{t,2} = 0.5y_{t,1}+0.4W_{N_{1}}y_{t,1}+0.6W_{N_{2}}y_{t,2}+0.9x_{t,2}+u_{t,2}
\end{aligned}
\end{equation}

where $y_{t,1}$ is a vector of half of the $N$ dependent variables observed at time $t$, and  $y_{t,2}$ is the other half.  When setting $W_{N_{1}}$ and $W_{N_{2}}$, we assume a
variable in $y_{t,1}$ is only spatially related with the unit ahead of it and the unit behind, likewise a variable in $y_{t,2}$. Both $W_{N_{1}}$ and $W_{N_{2}}$ are normalized by
rows. In Monte Carlo, we generate each element of $x_{t,1}$ and $x_{t,2}$ independently from $N(0,1)$, then each element of $u_{t,1}$ and $u_{t,2}$ independently from
$0.1N(0,1)$.\footnote{We have also experimented on $W_{N}$, $W_{N_{1}}$ and $W_{N_{2}}$ of other forms. In addition, we have looked into models with different coefficients,
$u_{t,1}\sim 0.1N(0,1)$ and  $u_{t,2}\sim 0.3N(0,1)$. The Monte Carlo results provide further evidence that two-stage VB method works well.}

For both (\ref{simuSAr})  and (\ref{liu_sa}), the sample sizes considered are $N=30\hspace{0.1cm}$ and $\hspace{0.1cm} T=20$,
$N=30\hspace{0.1cm}$ and $\hspace{0.1cm} T=80$,
$N=50\hspace{0.1cm}$ and $\hspace{0.1cm} T=30$,
$N=50\hspace{0.1cm}$ and $\hspace{0.1cm} T=100$,
$N=100\hspace{0.1cm}$ and $\hspace{0.1cm} T=50$, and
$N=100\hspace{0.1cm}$ and $\hspace{0.1cm} T=200$. For each case, we conduct 1000 Monte Carlo replications and use the changes in parameters instead of that of  ELBO to check
whether two-stage VB has converged.

Results of Monte Carlo simulations, which are relegated to Online Supplement to save space, provide strong evidence that the two-stage VB is able to recover the true parameters in
the data generating process, especially when $T\gg N$. When $T\ll N$, two-stage VB estimates have larger biases and larger empirical standard deviations, which can be reduced by
setting tighter priors. More important, there is clear evidence that the true spatial connections can be identified regardless of the length of the panel, long or short.

To give a flavour, tables 1-2 report the Monte Carlo results of parameter $\Lambda$  and $\tilde{\beta}$ for model  (\ref{simuSAr})  with $N=30$ and $T=80$ as well as $N=30$ and
$T=20$. Since a tablet with 30 columns is too big to fit into a page, we only report the mean and standard deviations, the latter in parenthesis, of the empirical distributions of
the $5\times 5$ sub-matrices in the four corners of $\Lambda$, which are associated with $y_{t,1},...,y_{t,5}$ and $y_{t,26},...,y_{t,30}$, and the first and last five elements in
$\tilde{\beta}$.

\begin{table}[h!]
\centering
\caption{Estimates of $\Lambda$ for model (\ref{simuSAr}), where $N=30$}
\begin{tiny}
\noindent\begin{tabular}{ccccccccccc}\hline\hline
	T=80	&		&		&		&		&	…	&		&		&		&		&		\\\hline\hline
		&	0.30	&	0.00	&	0.00	&	0.00	&	…	&	0.00	&	0.00	&	0.00	&	0.00	&	0.30	\\
 		&(	0.02	)&(	0.02	)&(	0.02	)&(	0.02	)&	…	&(	0.02	)&(	0.02	)&(	0.02	)&(	0.02	)&(	0.02	)\\\hline
	0.30	&		&	0.30	&	0.00	&	0.00	&	…	&	0.00	&	0.00	&	0.00	&	0.00	&	0.00	\\
(	0.02	)&&(	0.02	)&(	0.02	)&(	0.02	)&	…	&(	0.02	)&(	0.02	)&(	0.02	)&(	0.02	)&(	0.02	)\\\hline
	0.00	&	0.30	&		&	0.30	&	0.00	&	…	&	0.00	&	0.00	&	0.00	&	0.00	&	0.00	\\
(	0.02	)&(	0.02	)&&(	0.02	)&(	0.02	)&	…	&(	0.02	)&(	0.02	)&(	0.02	)&(	0.02	)&(	0.02	)\\\hline
	0.00	&	0.00	&	0.30	&		&	0.30	&	…	&	0.00	&	0.00	&	0.00	&	0.00	&	0.00	\\
(	0.02	)&(	0.02	)&(	0.02	)&&(	0.02	)&	…	&(	0.02	)&(	0.02	)&(	0.02	)&(	0.02	)&(	0.02	)\\\hline
	0.00	&	0.00	&	0.00	&	0.30	&		&	…	&	0.00	&	0.00	&	0.00	&	0.00	&	0.00	\\
(	0.02	)&(	0.02	)&(	0.02	)&(	0.01	)&&	…	&(	0.02	)&(	0.02	)&(	0.02	)&(	0.02	)&(	0.02	)\\\hline
	…	&	…	&	…	&	…	&	…	&	…	&	…	&	…	&	…	&	…	&	…	\\
	0.00	&	0.00	&	0.00	&	0.00	&	0.00	&	…	&		&	0.30	&	0.00	&	0.00	&	0.00	\\
(	0.02	)&(	0.02	)&(	0.02	)&(	0.02	)&(	0.02	)&	…	&		&(	0.02	)&(	0.02	)&(	0.02	)&(	0.02	)\\\hline
	0.00	&	0.00	&	0.00	&	0.00	&	0.00	&	…	&	0.30	&		&	0.30	&	0.00	&	0.00	\\
(	0.02	)&(	0.02	)&(	0.02	)&(	0.02	)&(	0.02	)&	…	&	(0.02	)&&(	0.02	)&(	0.02	)&(	0.02	)\\\hline
	0.00	&	0.00	&	0.00	&	0.00	&	0.00	&	…	&	0.00	&	0.30	&		&	0.30	&	0.00	\\
(	0.02	)&(	0.02	)&(	0.02	)&(	0.02	)&(	0.02	)&	…	&	(0.02	)&(	0.02	)&(		)&(	0.01	)&(	0.02	)\\\hline
	0.00	&	0.00	&	0.00	&	0.00	&	0.00	&	…	&	0.00	&	0.00	&	0.30	&		&	0.30	\\
(	0.02	)&(	0.02	)&(	0.02	)&(	0.02	)&(	0.02	)&	…	&	(0.02	)&(	0.02	)&(	0.01	)&&(	0.02	)\\\hline
	0.30	&	0.00	&	0.00	&	0.00	&	0.00	&	…	&(	0.00	&	0.00	&	0.00	&	0.30	&		\\
(	0.02	)&(	0.02	)&(	0.02	)&(	0.02	)&(	0.02	)&	…	&	0.02	)&(	0.02	)&(	0.02	)&(	0.02	)&\\\hline\hline
	T=20	&		&		&		&		&	…	&		&		&		&		&		\\\hline\hline
		&	0.27	&	0.04	&	0.00	&	0.00	&	…	&	0.00	&	0.00	&	0.00	&	0.03	&	0.27	\\
&(	0.08	)&(	0.08	)&(	0.08	)&(	0.08	)&	…	&(	0.08	)&(	0.08	)&(	0.08	)&(	0.09	)&(	0.08	)\\\hline
	0.27	&		&	0.27	&	0.04	&	0.00	&	…	&	0.00	&	0.00	&	0.00	&	0.00	&	0.03	\\
(	0.08	)&&(	0.08	)&(	0.09	)&(	0.08	)&	…	&(	0.08	)&(	0.08	)&(	0.08	)&(	0.08	)&(	0.09	)\\\hline
	0.04	&	0.27	&		&	0.27	&	0.03	&	…	&	0.00	&	0.00	&	0.00	&	0.00	&	0.00	\\
(	0.09	)&(	0.08	)&&(	0.08	)&(	0.09	)&	…	&(	0.08	)&(	0.08	)&(	0.08	)&(	0.09	)&(	0.08	)\\\hline
	0.00	&	0.03	&	0.27	&		&	0.27	&	…	&	0.00	&	0.00	&	0.00	&	0.00	&	0.00	\\
(	0.08	)&(	0.08	)&(	0.08	)&&(	0.08	)&	…	&(	0.08	)&(	0.09	)&(	0.08	)&(	0.08	)&(	0.08	)\\\hline
	0.00	&	0.00	&	0.03	&	0.27	&		&	…	&	0.00	&	0.00	&	0.00	&	0.00	&	0.00	\\
(	0.08	)&(	0.08	)&(	0.08	)&(	0.08	)&&	…	&(	0.08	)&(	0.08	)&(	0.08	)&(	0.08	)&(	0.08	)\\\hline
	…	&	…	&	…	&	…	&	…	&	…	&	…	&	…	&	…	&	…	&	…	\\
	0.00	&	0.00	&	0.00	&	0.00	&	0.00	&	…	&		&	0.27	&	0.03	&	0.00	&	0.00	\\
(	0.08	)&(	0.08	)&(	0.08	)&(	0.08	)&(	0.08	)&	…	&&(	0.08	)&(	0.08	)&(	0.08	)&(	0.08	)\\\hline
	0.00	&	0.00	&	0.00	&	0.00	&	0.00	&	…	&	0.27	&		&	0.27	&	0.04	&	0.00	\\
(	0.08	)&(	0.08	)&(	0.08	)&(	0.09	)&(	0.08	)&	…	&(	0.07	)&&(	0.08	)&(	0.09	)&(	0.08	)\\\hline
	0.00	&	0.00	&	0.00	&	0.00	&	0.00	&	…	&	0.03	&	0.27	&		&	0.27	&	0.04	\\
(	0.09	)&(	0.08	)&(	0.08	)&(	0.09	)&(	0.09	)&	…	&(	0.08	)&(	0.08	)& &(	0.08	)&(	0.08	)\\\hline
	0.04	&	0.00	&	0.00	&	0.00	&	0.00	&	…	&	0.00	&	0.04	&	0.27	&		&	0.27	\\
(	0.09	)&(	0.08	)&(	0.08	)&(	0.08	)&(	0.08	)&	…	&(	0.08	)&(	0.08	)&(	0.08	)&&(	0.08	)\\\hline
	0.27	&	0.03	&	0.00	&	0.00	&	0.00	&	…	&	-0.01	&	0.00	&	0.04	&	0.27	&		\\
(	0.08	)&(	0.08	)&(	0.08	)&(	0.08	)&(	0.08	)&	…	&(	0.08	)&(	0.09	)&(	0.08	)&(	0.08	)&\\\hline
 \hline
\end{tabular}
\end{tiny}
\label{table:1}
\end{table}

\begin{table}[h!]
\centering
\caption{Estimates of $\tilde{\beta}$ for model (\ref{simuSAr}), where $N=30$}
\begin{tiny}
\noindent\begin{tabular}{ccccccccccc}\hline\hline
T=80	&		&		&		&		&		&		&		&		&		&		\\\hline\hline
0.90	&	0.90	&	0.90	&	0.90	&	0.90	&...		&	0.90	&	0.90	&	0.90	&	0.90	&	0.90	\\
(0.02	)&(	0.02	)&(	0.01	)&(	0.02	)&(	0.02	)&...&(	0.02	)&(	0.02	)&(	0.02	)&(	0.02	)&(	0.02	)\\\hline\hline
																					
T=20	&		&		&		&		&		&		&		&		&		&		\\\hline\hline
0.65	&	0.66	&	0.65	&	0.66	&	0.65	&	...	&	0.65	&	0.65	&	0.65	&	0.65	&	0.65	\\
(0.08	)&(	0.07	)&(	0.07	)&(	0.07	)&(	0.08	)&...&(	0.07	)&(	0.08	)&(	0.07	)&(	0.08	)&(	0.08	)\\\hline
 \hline
\end{tabular}
\end{tiny}
\label{table:2}
\end{table}

We use high-performance computing (HPC) services for estimation. All computations are done using 1 compute node and 4 processor core. On average, each Monte Carlo replication for
model (\ref{simuSAr}) where $N=30$ and $T=80$ takes about 20 seconds, while for the same model, when $N=30$ and $T=20$ it takes about 60 seconds. For model (\ref{liu_sa}), it again
takes 20 seconds when $N=30$ and $T=80$. However, estimating model (\ref{liu_sa}) takes 3 minutes when $N=30$ and $T=20$.  The same pattern can be observed when estimating models
where $N=50$ and $N=100$, that is: two-stage VB takes much longer to converge when $N\gg T$, and the larger the number of nonzero true parameters, the longer the estimation takes.
To give a flavour of how fast two-stage VB converges, we would like to mention that, when allowing for parallel computing, each Monte Carlo replication of model (\ref{simuSAr})
where $N=100$ and $T=50$ takes about 20 minutes, and that number reduces to 5 minutes for the same model where $N=100$ and $T=200$.

\section{Empirical Applications}
The relationship between sovereign bond ratings and spreads in eurozone countries is of great interest to researchers and policy makers alike, as explained in papers such as Gibson
et al. (2021) and Hall et al. (2022). One of the important questions posed is whether rate changes of the southern euro area countries and their northern counterparts have
different impacts on a member country, especially during the European debt crisis, also known as euro area crises, that started in late 2009 and lingered on well into 2015.
Equally, it is important to know whether rate changes of a country group affect all the other countries  differently. To answer those questions, we apply the two-stage VB approach
on a sample containing five southern euro area countries -- Spain, Greece, Ireland, Italy and Portugal -- and five northern euro area countries -- Austria, Belgium, France and
Netherlands.

We use monthly data of Gibson et al. (2021).  Sovereign rating is denoted by Rate, which is the combined sovereign ratings given by Standard and Poor's, Fitch and Moody's then
transformed into numerical scale with `triple A' having value 1 and `selected default' having the value 22. Spread is defined as the difference between the yield on 10-year
government bond of a country and that of Germany. Hence a rise in the rate or spread implies a worsening situation. Following Gibson et al. (2017), we use the following variables
that capture the economic and political fundamentals as control variables:
GDPgrowth, which is  the real GDP growth rate; News, which is fiscal news constructed using European Commision forecasts; $\frac{Debt}{GDP}$, which is the ratio of government debt
to GDP;  $\frac{CA}{GDP}$, which is the ratio of current account balance to GDP;  $\frac{P}{P^{*}}$, which is the ratio of a country's harmonised consumer price to that of Germany;
and Pol, which is the index of political uncertainty reflecting the climate for foreign investors and political uncertainty. The monthly data runs from January 2000 to April 2019.
For brevity, we report the data sources in  the Appendix.

Taking account of the feedback loop between sovereign bond ratings and sovereign spreads, our dynamic model takes the following form:
 \begin{equation}\label{euro_model}
  \begin{aligned}
 Rate_{it} = & c_{11,i}+c_{12,i}*Spread_{it}+c_{13,i}*GDPgrowth_{it}+c_{14,i}*\frac{Debt_{it}}{GDP_{it}} +c_{15,i}*News+\\
&c_{16,i}*\frac{P}{P^{*}}_{it}+c_{17,i}*Rate_{i(t-1)}+W_{i}^{rate}*Rate_{t}+\varepsilon_{it}^{rate}\\
Spread_{it} = &c_{21,i} +c_{22,i}*Rate_{it}+c_{23,i}*GDPgrowth_{it}+c_{24,i}*\frac{Debt_{it}}{GDP_{it}} +c_{25,i}*Pol_{it}+\\
&c_{26,i}*\frac{CA_{it}}{GDP_{it}}+c_{27,i}*Spread_{i(t-1)}+W_{i}^{sp}*Spread_{t}+\varepsilon_{it}^{sp}\\
\end{aligned}
\end{equation}
where $W_{i}^{rate}$ and $W_{i}^{spread}$ are the  $i^{th} $ row of the $10\times 10$ spatial weights matrices $W^{rate}$ and $W^{spread}$, respectively. Note that $W^{rate}$ and
$W^{spread}$ are both with zero diagonals and the rest of the elements to be estimated.

Model (\ref{euro_model}) can be rewritten as
\begin{equation}\label{euro_modellalala}
\begin{tiny}
\left[
    \begin{array}{c}
      Rate_{t} \\
      Spread_{t} \\
    \end{array}\right]
    = \left[
       \begin{array}{cc}
         I_{10}-W^{rate} & -\mathbf{c}_{12} \\
         -\mathbf{c}_{22} & I_{10}-W^{spread} \\
       \end{array}
     \right]^{-1}
    \left( \left[
       \begin{array}{cc}
         \mathbf{c}_{17} & 0 \\
         0 & \mathbf{c}_{27} \\
       \end{array}
     \right]
     \left[
    \begin{array}{c}
      Rate_{t-1} \\
      Spread_{t-1} \\
    \end{array}
    \right]+A\mathbf{x}+\left[
                          \begin{array}{c}
                            \varepsilon_{t}^{rate} \\
                            \varepsilon_{t}^{spread} \\
                          \end{array}\right]
                        \right)
\end{tiny}
\end{equation}
where $Rate_{t}=(Rate_{1t}, ...,Rate_{10t})'$, $Spread_{t}=(Spread_{1t}, ...,Spread_{10t})'$, $ \varepsilon_{t}^{rate}=( \varepsilon_{1t}^{rate},..., \varepsilon_{10t}^{rate})'$,
$\varepsilon_{t}^{spread}=(\varepsilon_{1t}^{spread},...,\varepsilon_{10t}^{spread})'$, $\mathbf{c}_{12}=diag(\mathbf{c}_{12,1},..., \mathbf{c}_{12,10})$,
$\mathbf{c}_{22}=diag(\mathbf{c}_{22,1},..., \mathbf{c}_{22,10})$,  $\mathbf{c}_{17}=diag(\mathbf{c}_{17,1},..., \mathbf{c}_{17,10})$,  $\mathbf{c}_{27}=diag(\mathbf{c}_{27,1},...,
\mathbf{c}_{27,10})$, $\mathbf{x}$ is the vector containing all the exogenous variables, and $A$ is the matrix containing their respective parameters. In the spirit of Debarsy et
al. (2012), we use equation (\ref{euro_modellalala}) to calculate the impulse response functions and evaluate how rate and spread changes in one country spill over to the other
countries.

We use a rolling window of 24 months to trace how the spillovers fluctuate over time. For the $i^{th}$ country, the average rating spillovers from the south is computed by taking
the mean of the cumulative impacts of a 1 notch value increase in ratings of southern euro area countries; and the average rating spillovers from the north is computed by the
average of the cumulative impacts of a 1 notch value increase in northern euro area countries. In both cases, the impact responses to a shock of own country are excluded. In the
same fashion, we calculate the average spread spillovers from the south and the north. We set the amount of shock to the spread to be 1 basis point.

\begin{figure}[!htbp]
    \includegraphics[width=.50\textwidth]{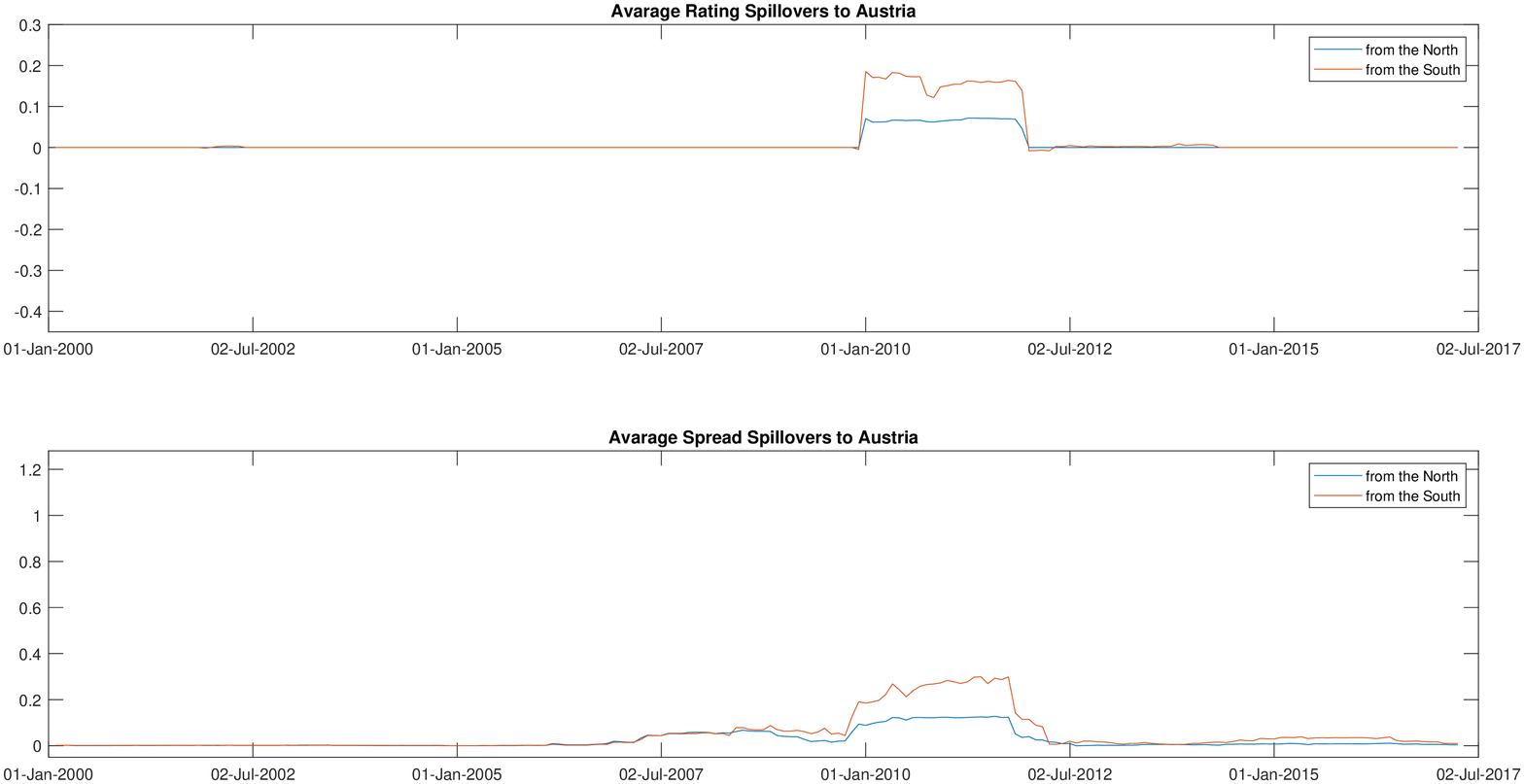}\hfill
    \includegraphics[width=.50\textwidth]{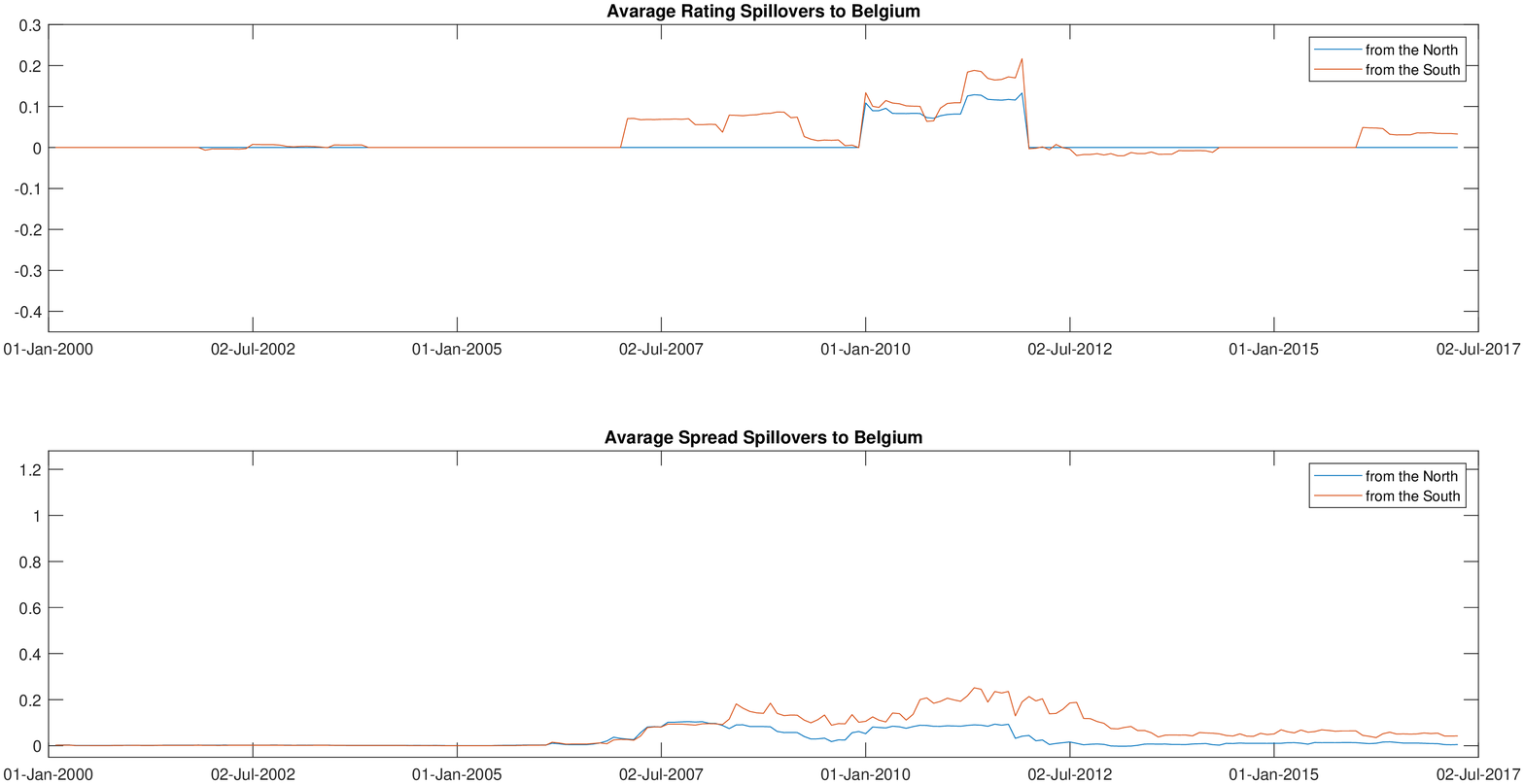}\hfill
    \\[\smallskipamount]
    \includegraphics[width=.50\textwidth]{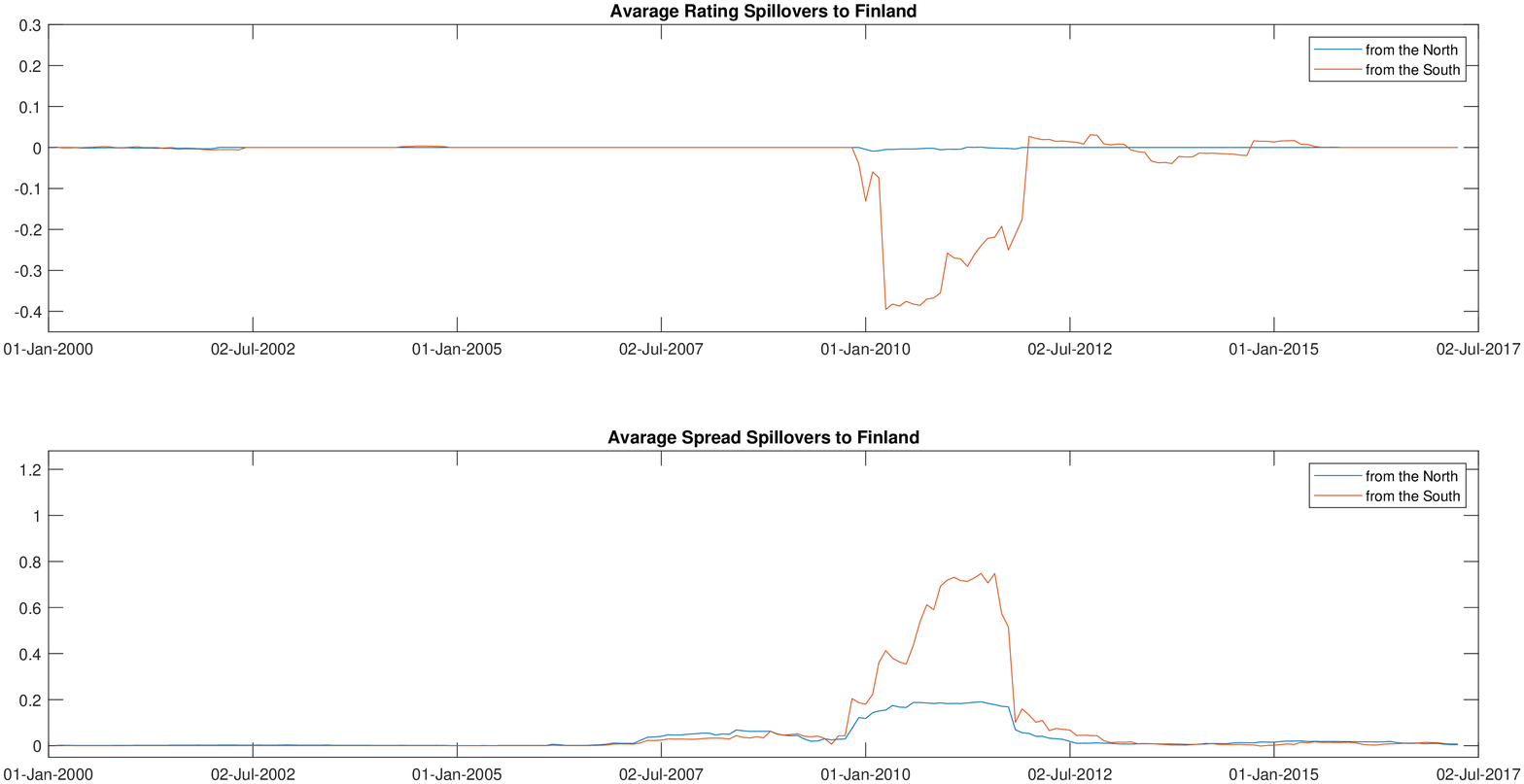}\hfill
    \includegraphics[width=.50\textwidth]{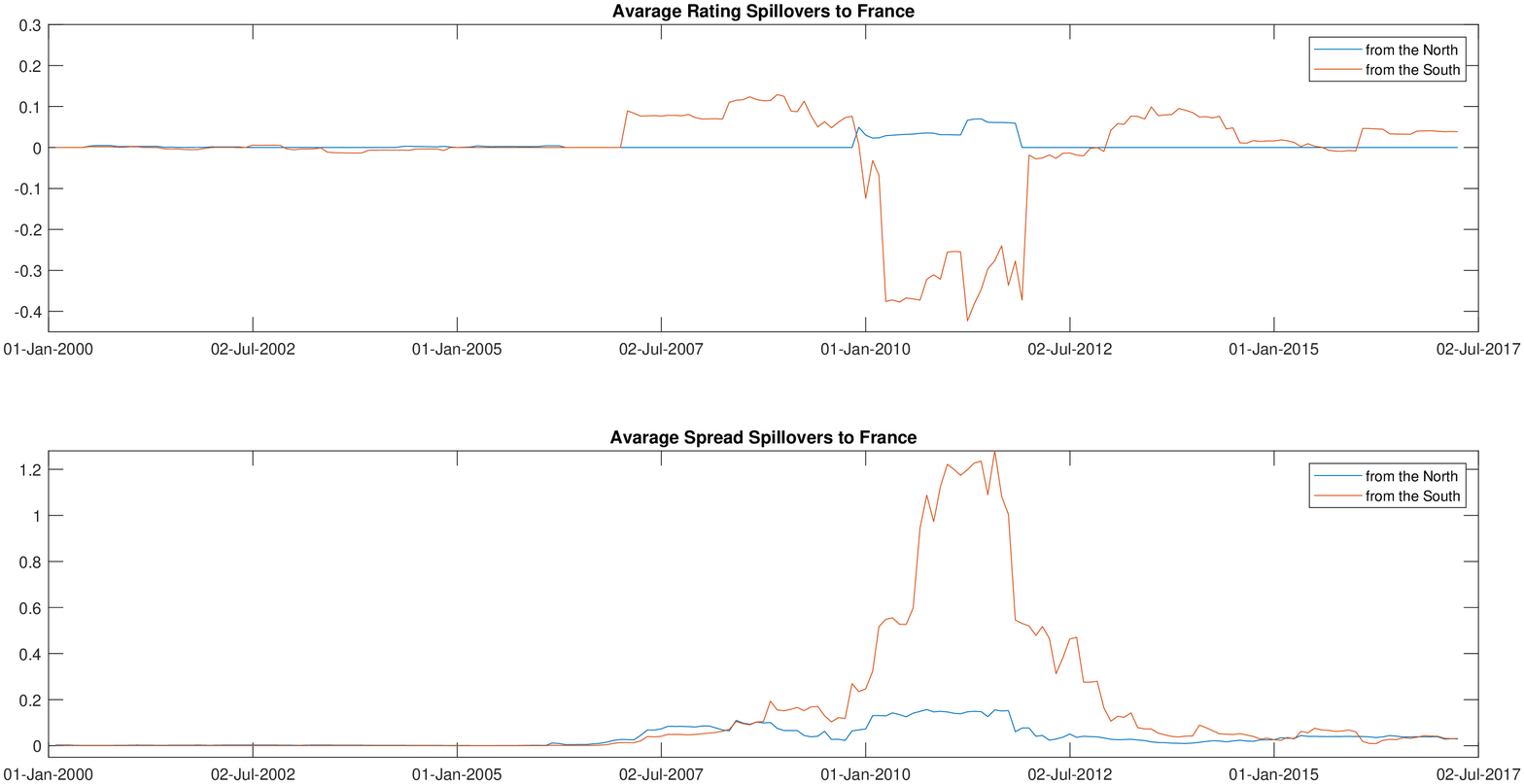}
  \\[\smallskipamount]
   \includegraphics[width=.50\textwidth]{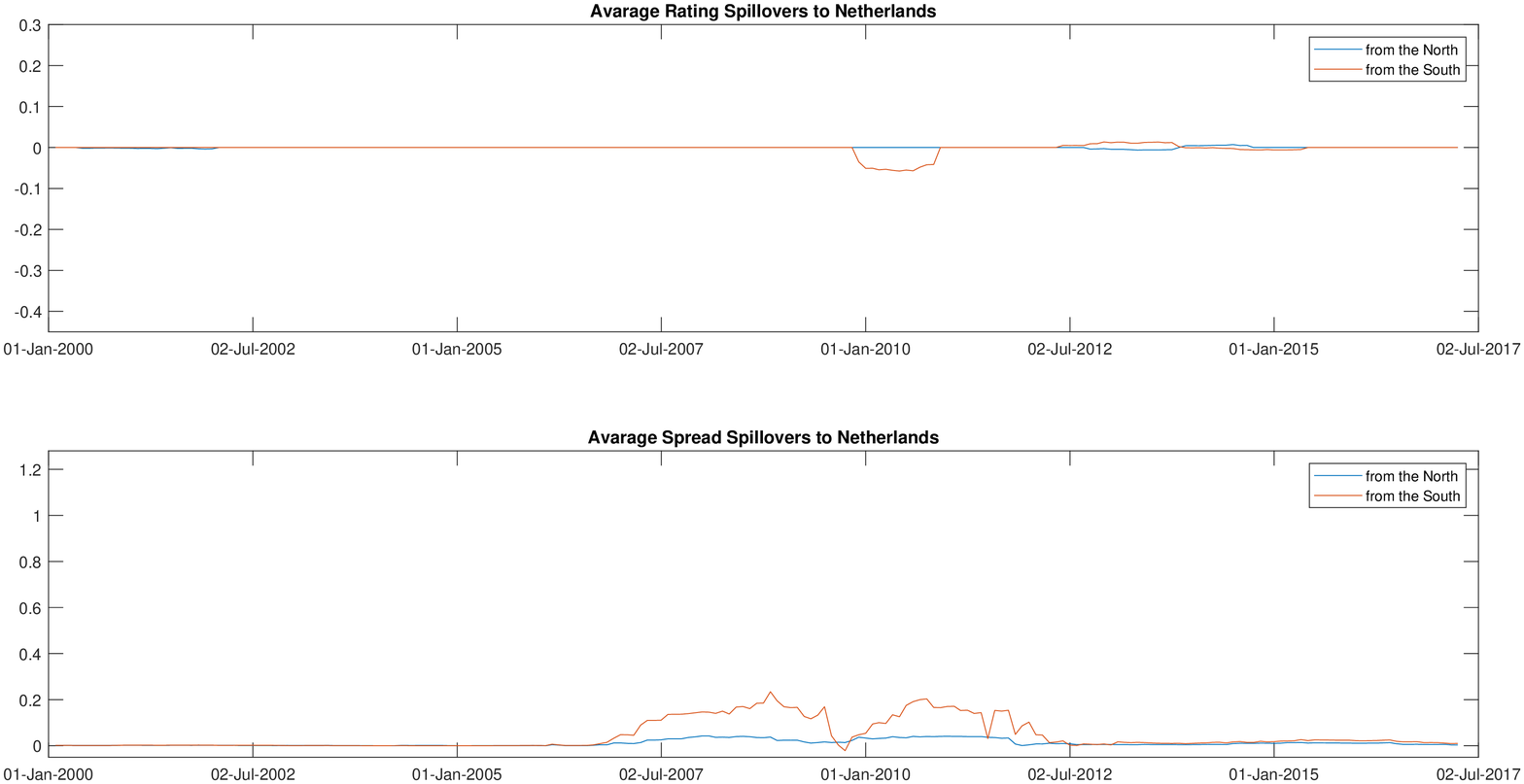}\hfill
    \includegraphics[width=.50\textwidth]{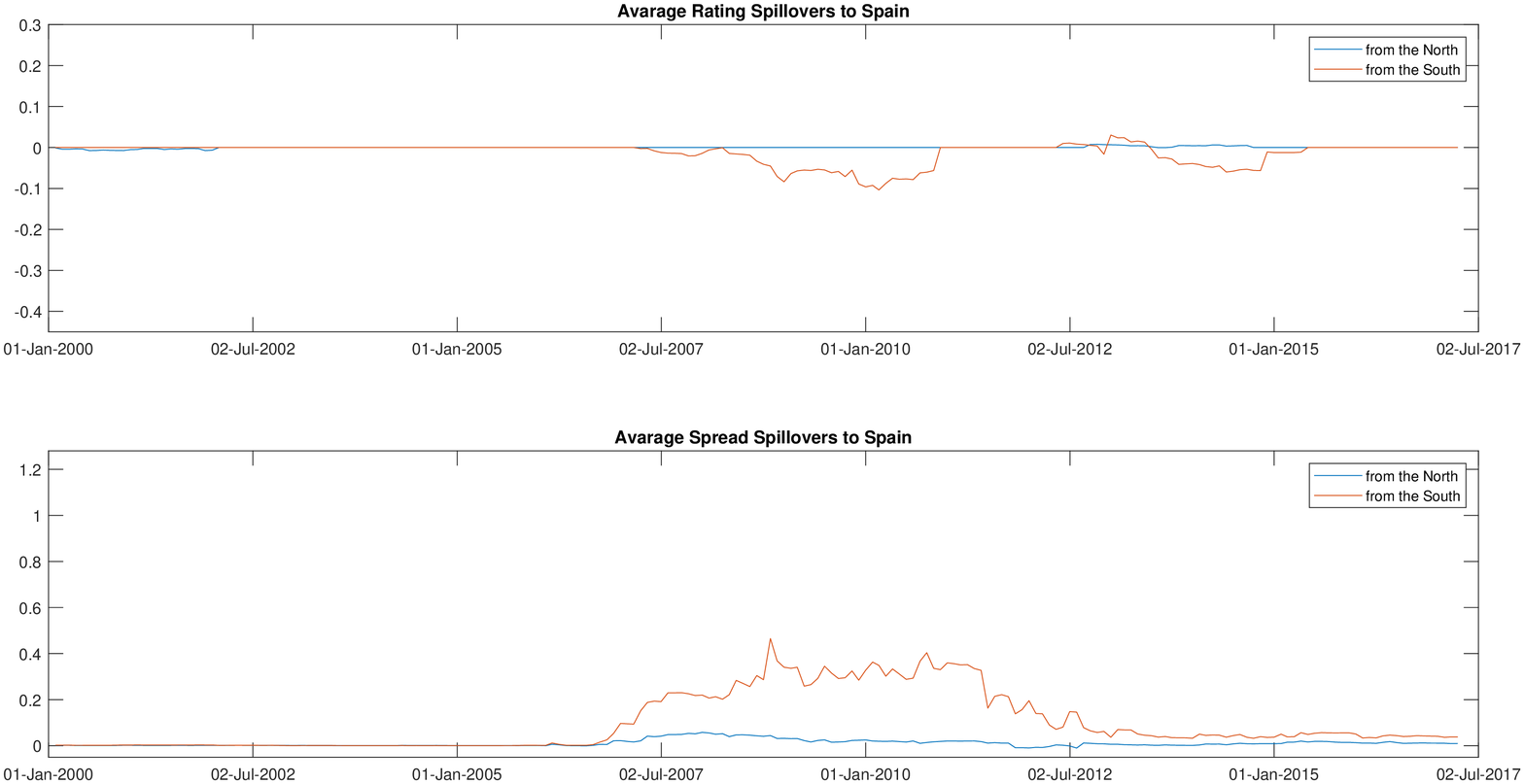}
  \\[\smallskipamount]
   \includegraphics[width=.50\textwidth]{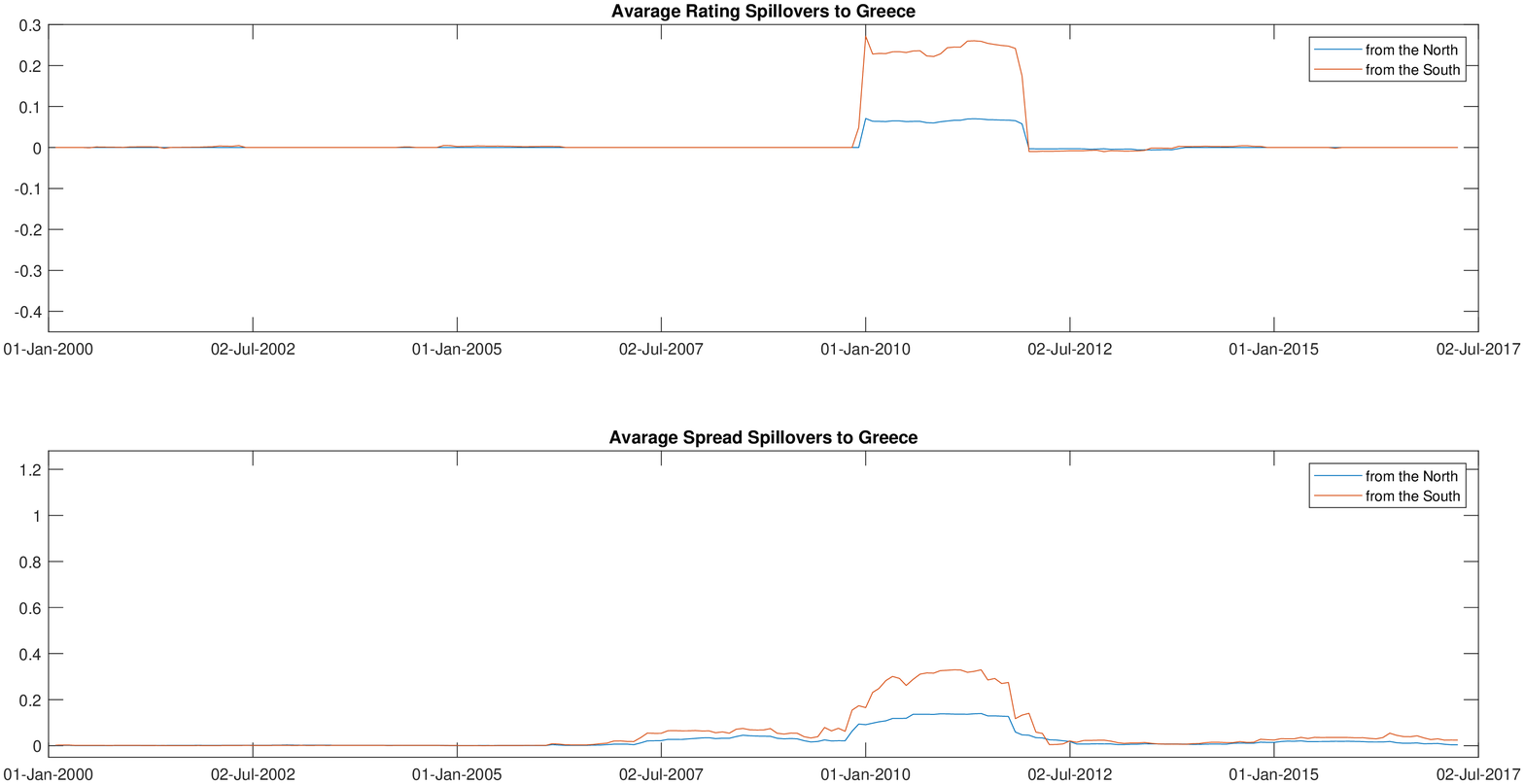}\hfill
    \includegraphics[width=.50\textwidth]{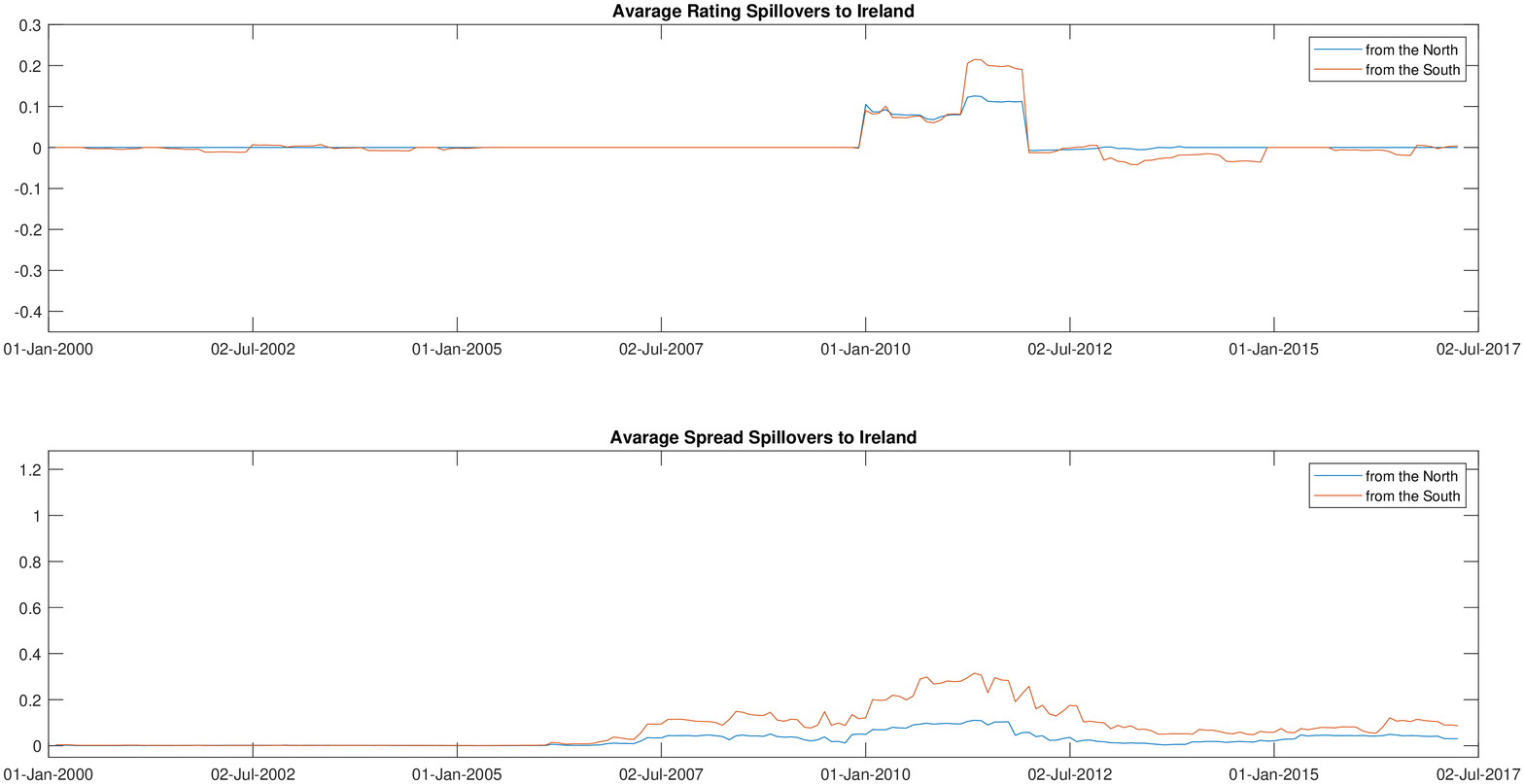}
  \\[\smallskipamount]
   \includegraphics[width=.50\textwidth]{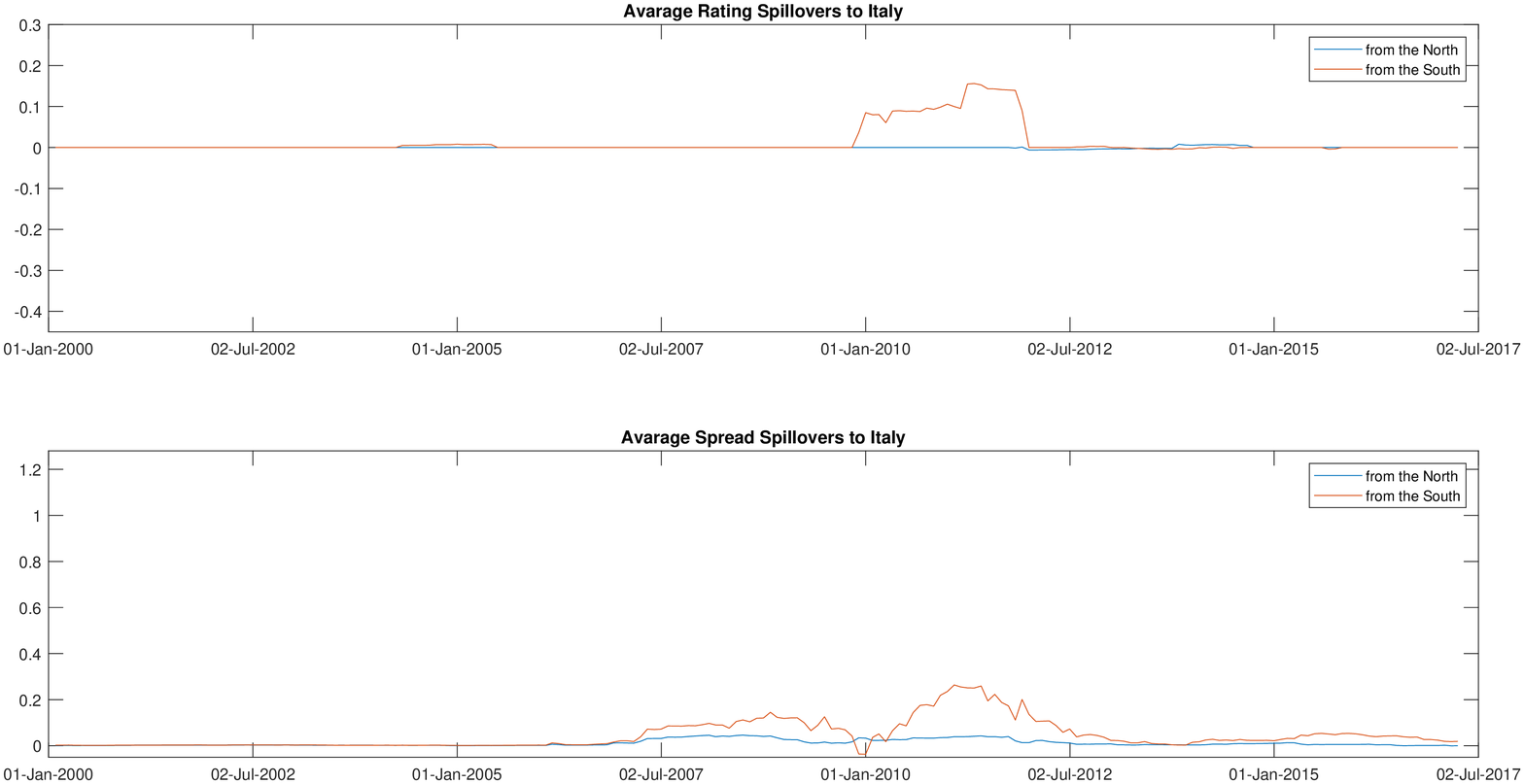}\hfill
    \includegraphics[width=.50\textwidth]{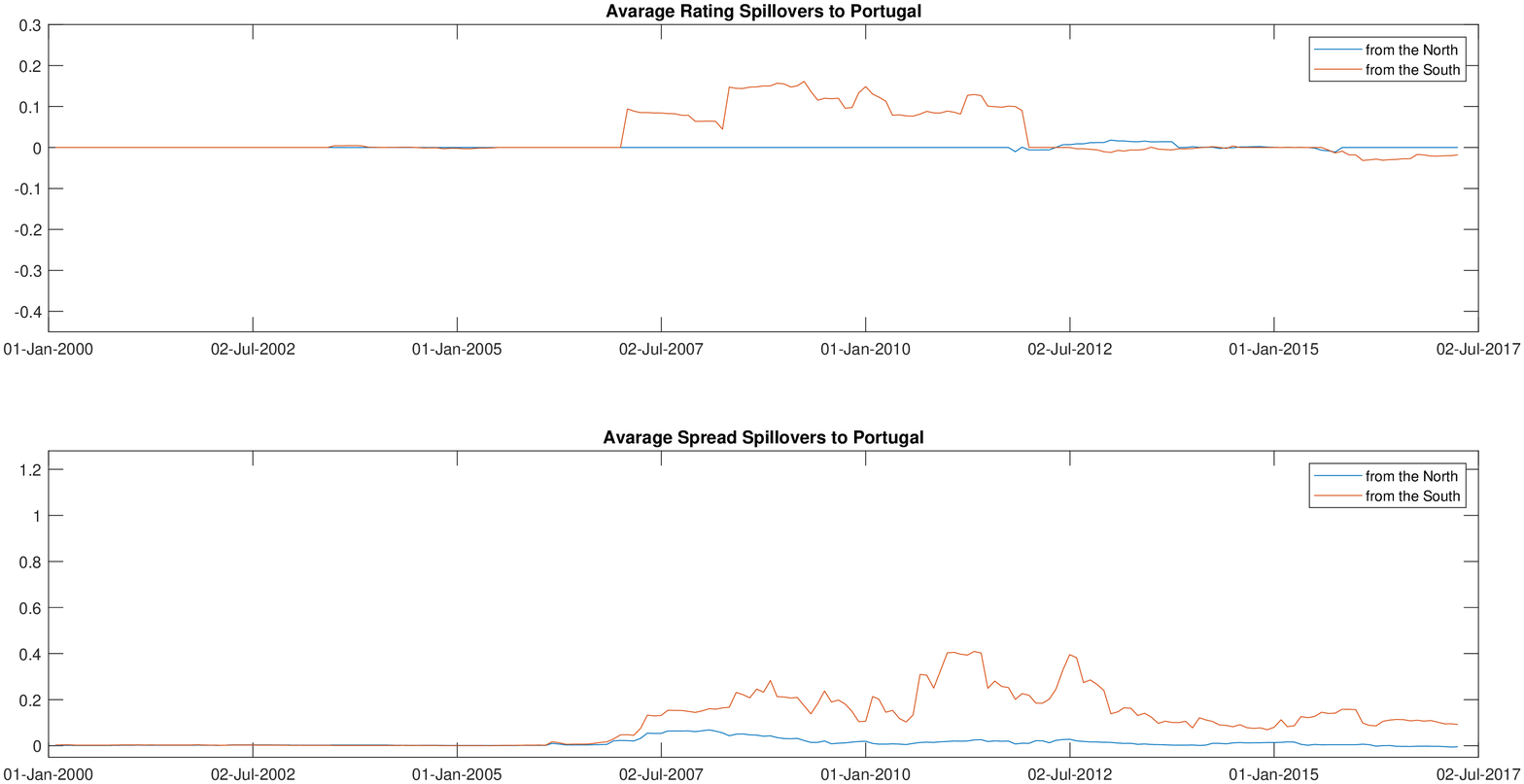}
  \\[\smallskipamount]
    \caption{Average spillovers}\label{fig:2_yr}
\end{figure}

Figure 1 presents the average spillovers from the southern euro area countries and those from the north, where spillovers are measured by the cumulative impacts of 5
years.\footnote{The impulse response functions usually converge to zero after 2 to 3 months} Two salient patterns leap out. First, spatial spillovers between countries only become
apparent after the financial crises in 2007. Before that, the spillovers between countries are barely minimum. Second, shocks to a country's spread tend to always increase other
countries' spreads, but the impacts of shocks to sovereign ratings are mixed.

Let us start with the impacts of rating changes in a northern euro area country. We can observe that shocks to rates of the southern countries tends to improve the sovereign
ratings of France, Finland and Netherland (causing the values of their rates to decrease), but worsen the ratings of Austria and Belgium (causing their rates to increase),
especially during the period of euro area crises. For a northern country, shocks to rates of other northern countries, by contrast, tend to slightly worsen its sovereign ratings.

For a southern country, apart from Greece and Ireland, shocks to rates of the northern countries tend to have little cumulative impacts on its ratings. For Greece and Ireland,
especially in the euro area crises, positive shocks to northern countries' ratings will cause their ratings to deteriorate as well. For all the southern countries, shocks that
worsen other southern countries' ratings will worsen their own ratings too, in a magnitude that is much higher than the impacts of the same amount of shocks to northern countries'
ratings.

For all countries, an increase in the spreads of northern countries is always followed by an increase in their own spreads, making it more costly for the country to borrow, same as
the impacts of an increase in the spreads of southern countries. But the latter's impacts are much larger than the former's in all cases. In euro area crises, the spreads of
France, Finland and Spain increased a lot after shocks to southern countries' spreads.

\begin{figure}[!htbp]
    \includegraphics[width=.50\textwidth]{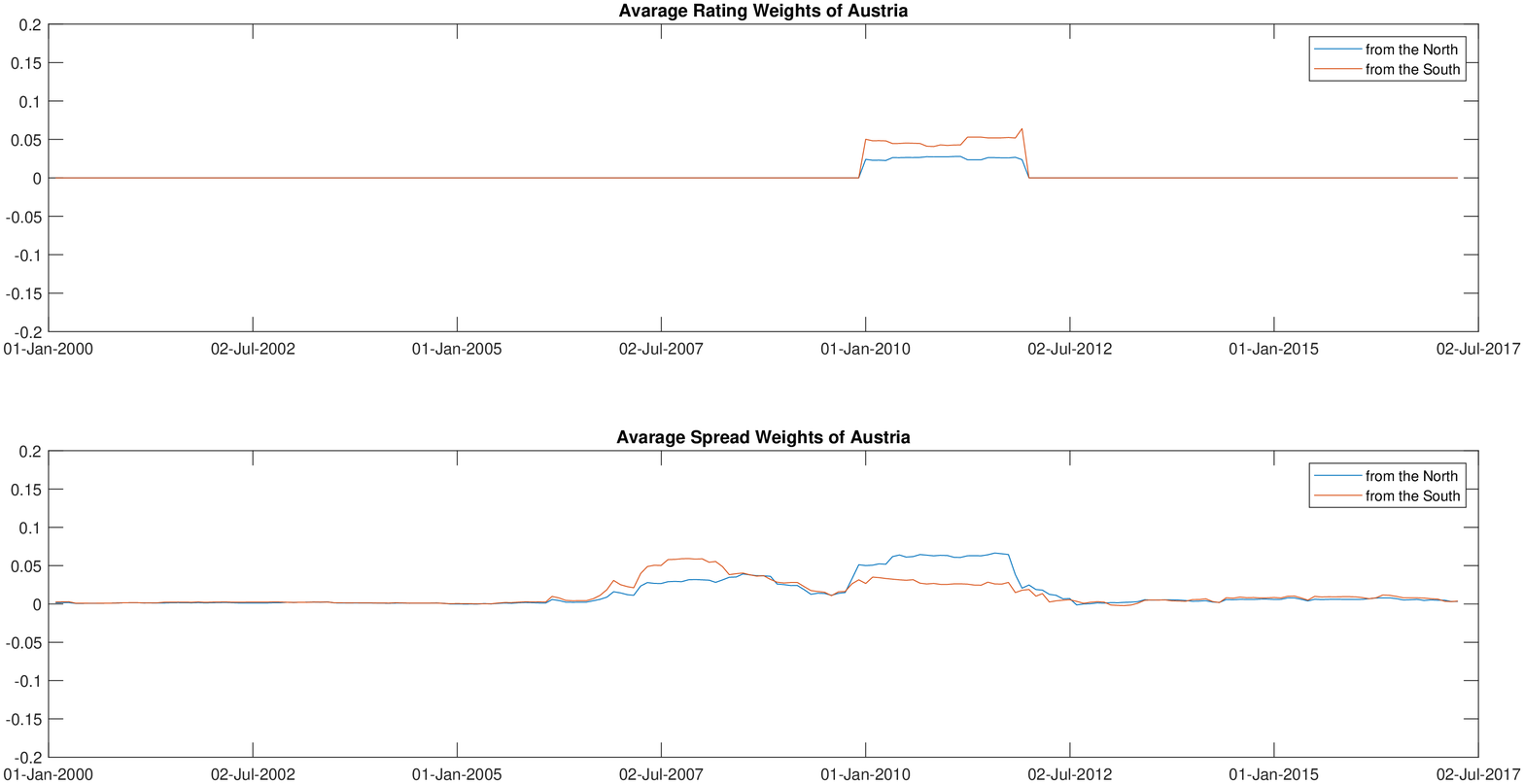}\hfill
    \includegraphics[width=.50\textwidth]{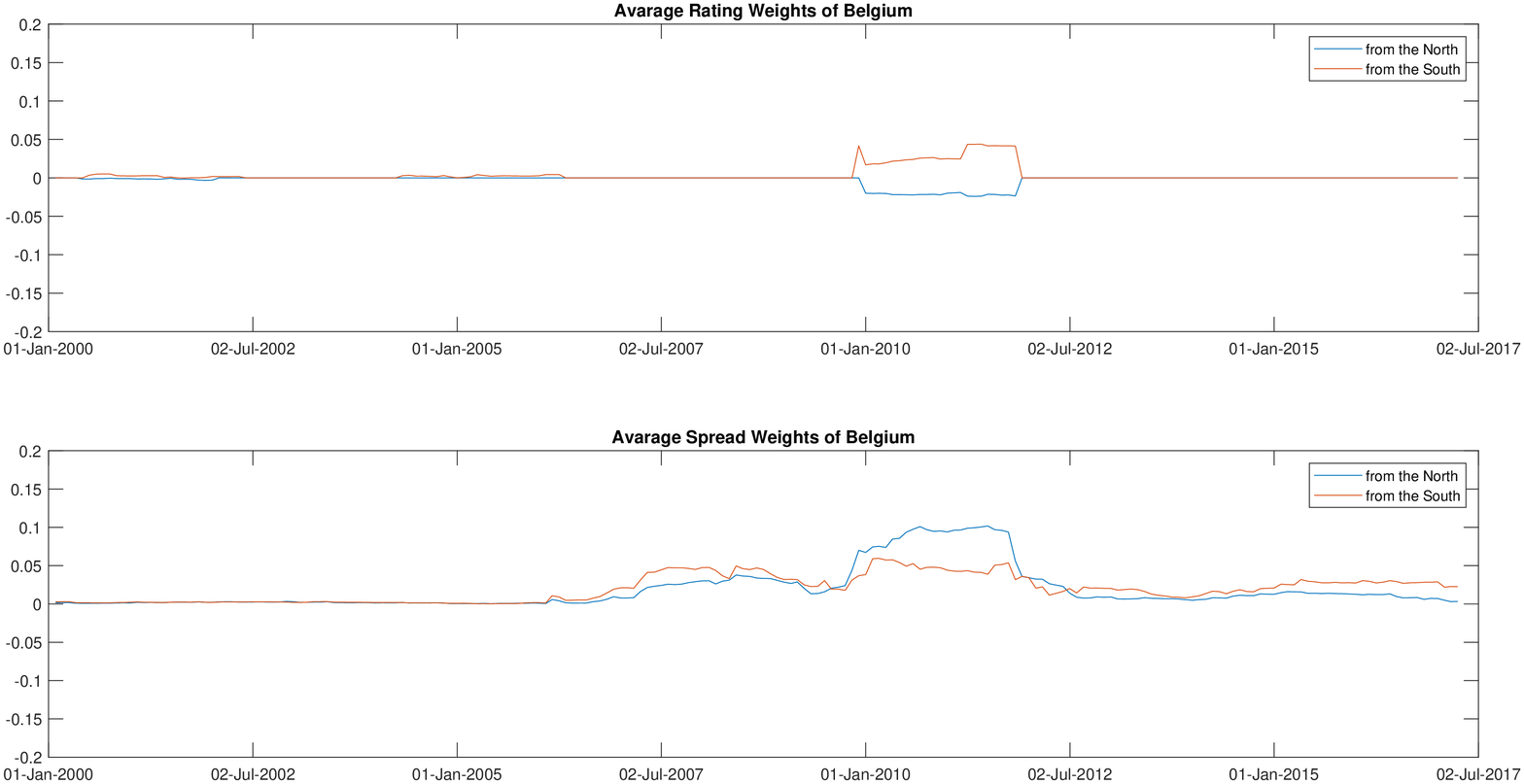}\hfill
    \\[\smallskipamount]
    \includegraphics[width=.50\textwidth]{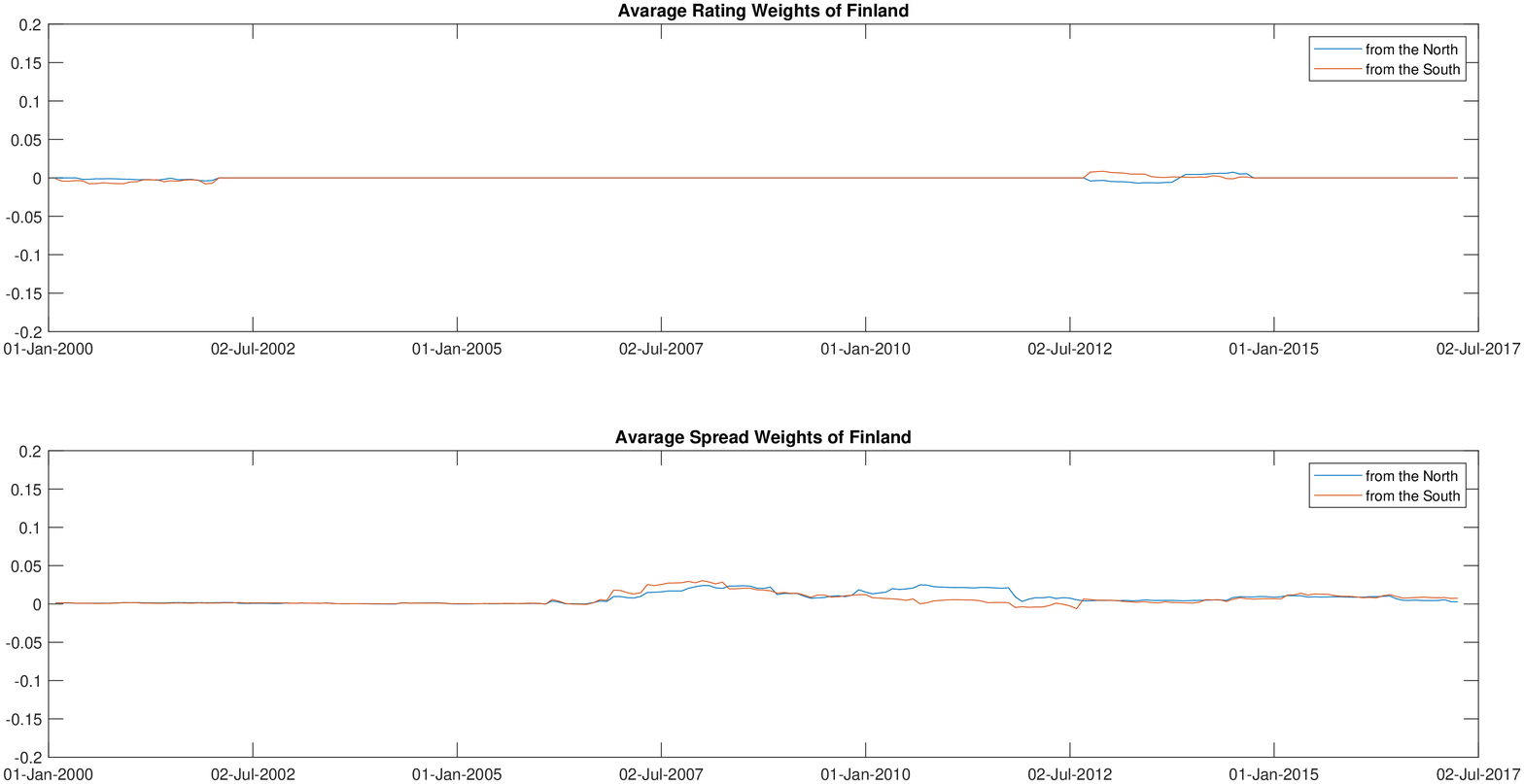}\hfill
    \includegraphics[width=.50\textwidth]{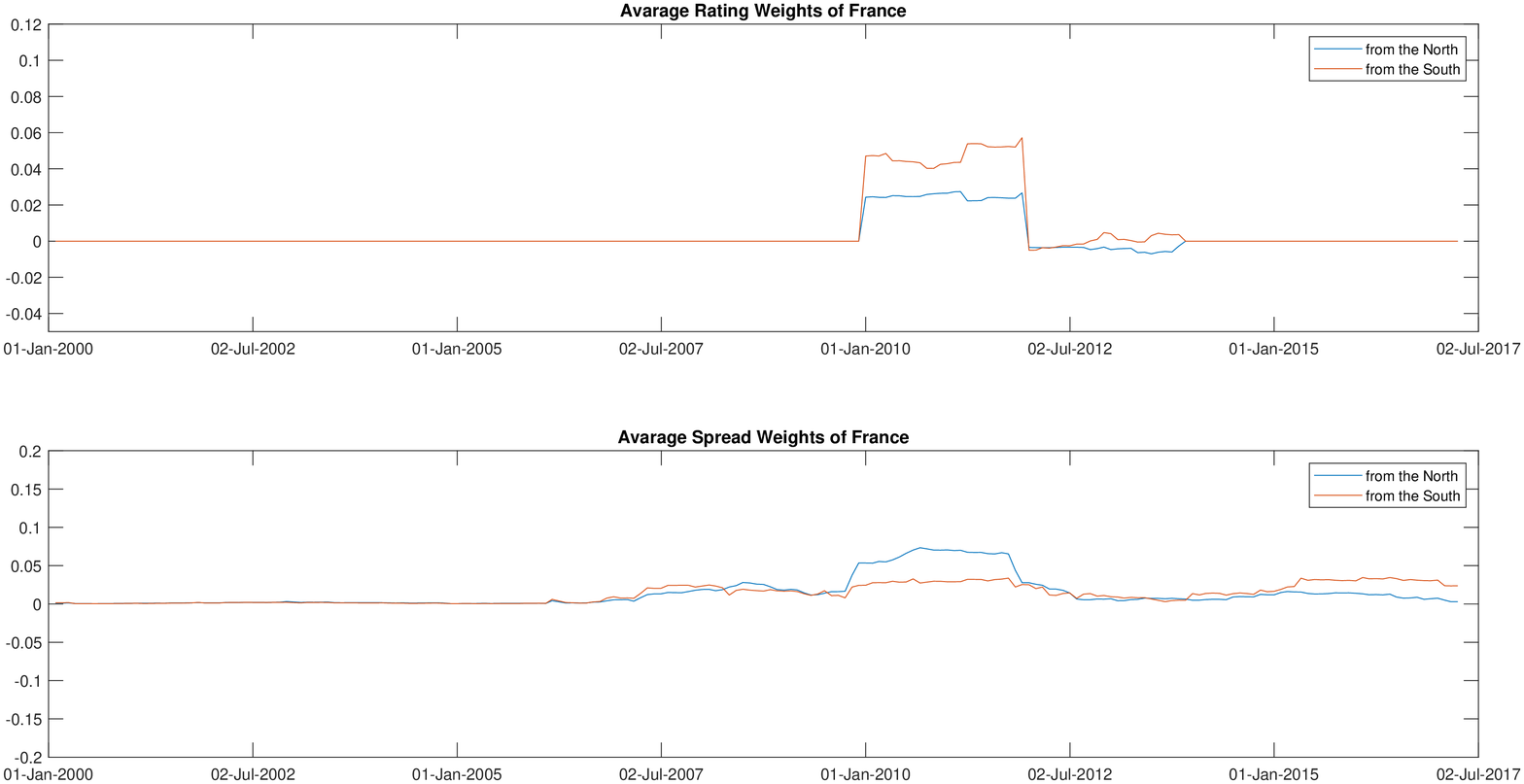}
  \\[\smallskipamount]
   \includegraphics[width=.50\textwidth]{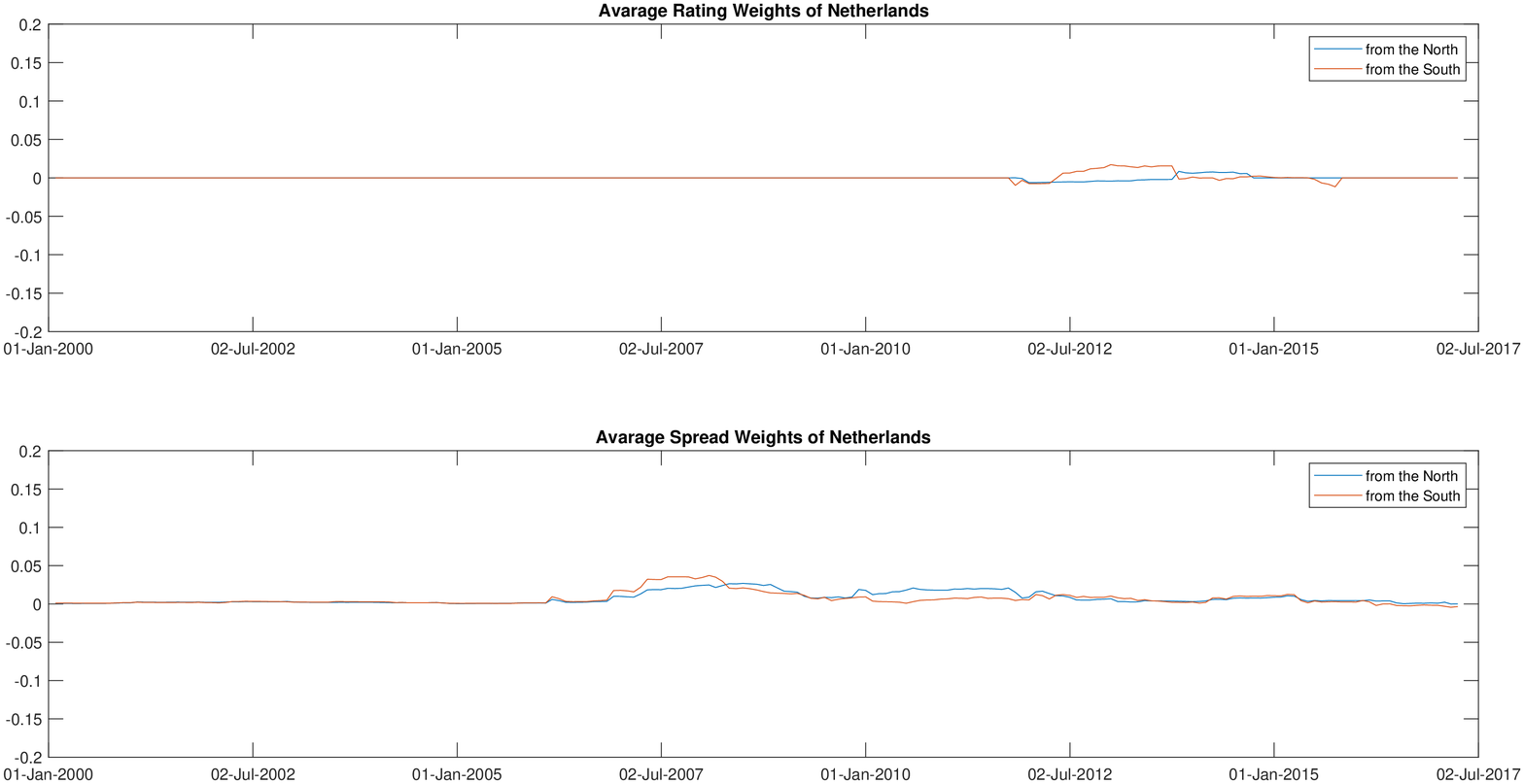}\hfill
    \includegraphics[width=.50\textwidth]{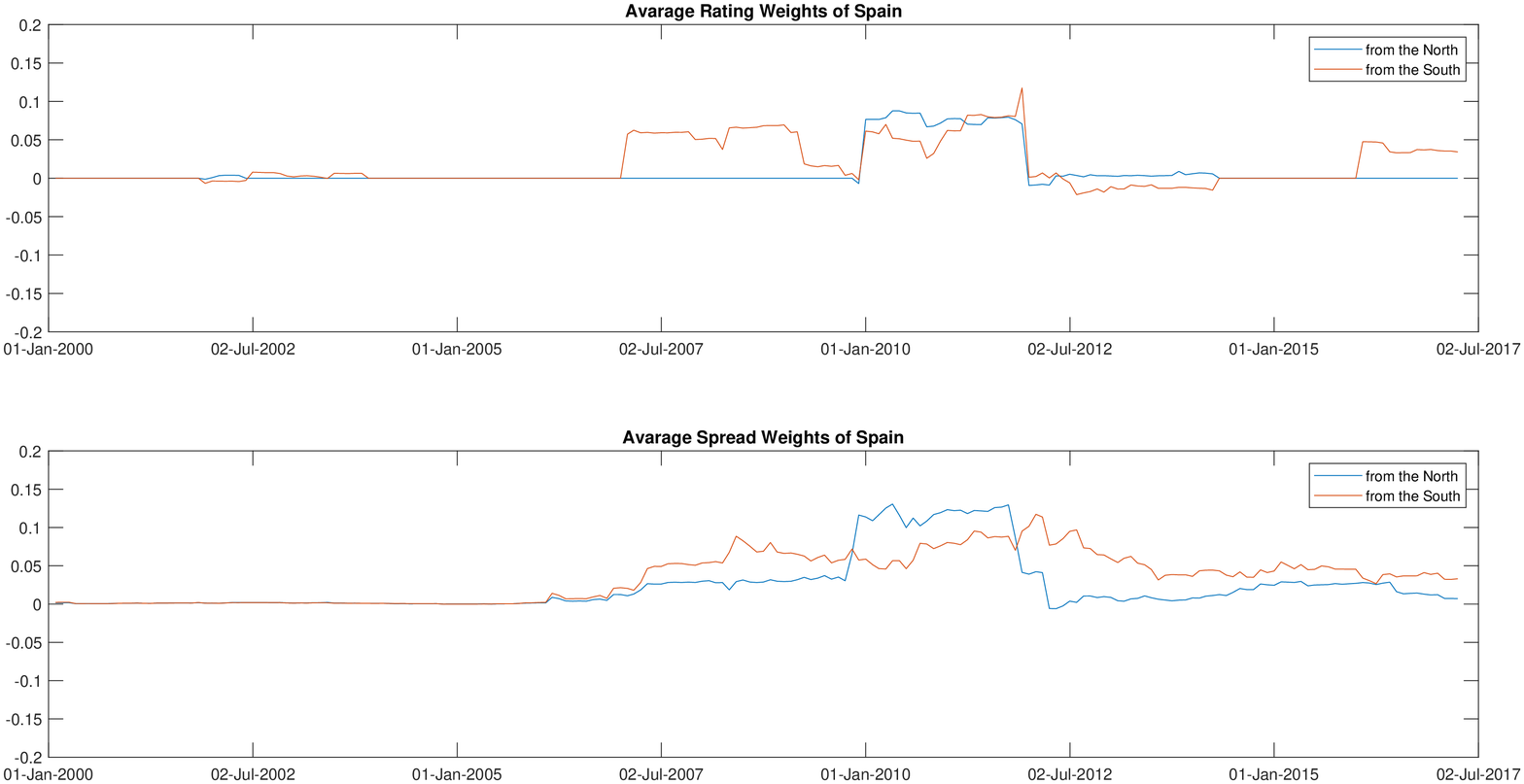}
  \\[\smallskipamount]
   \includegraphics[width=.50\textwidth]{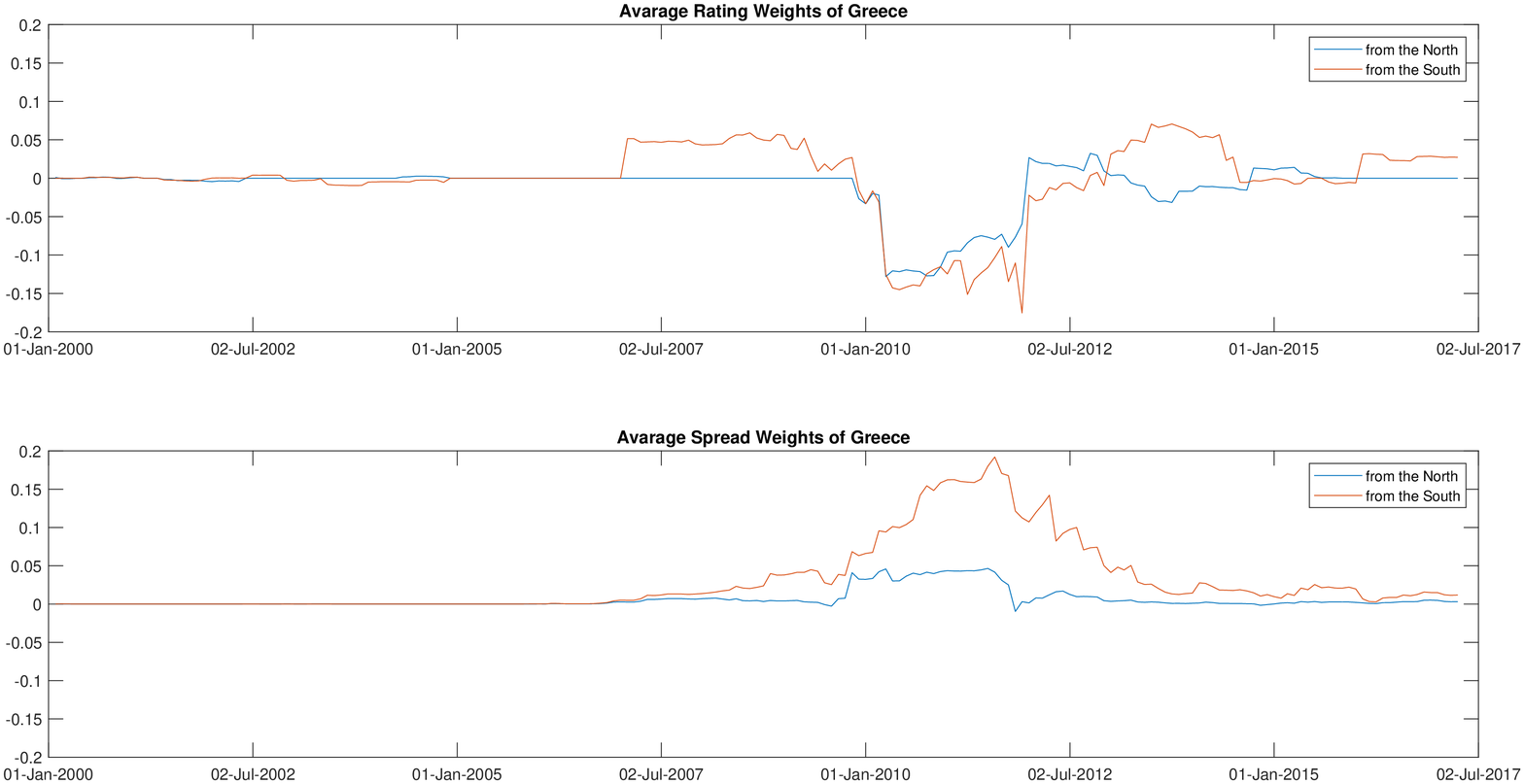}\hfill
    \includegraphics[width=.50\textwidth]{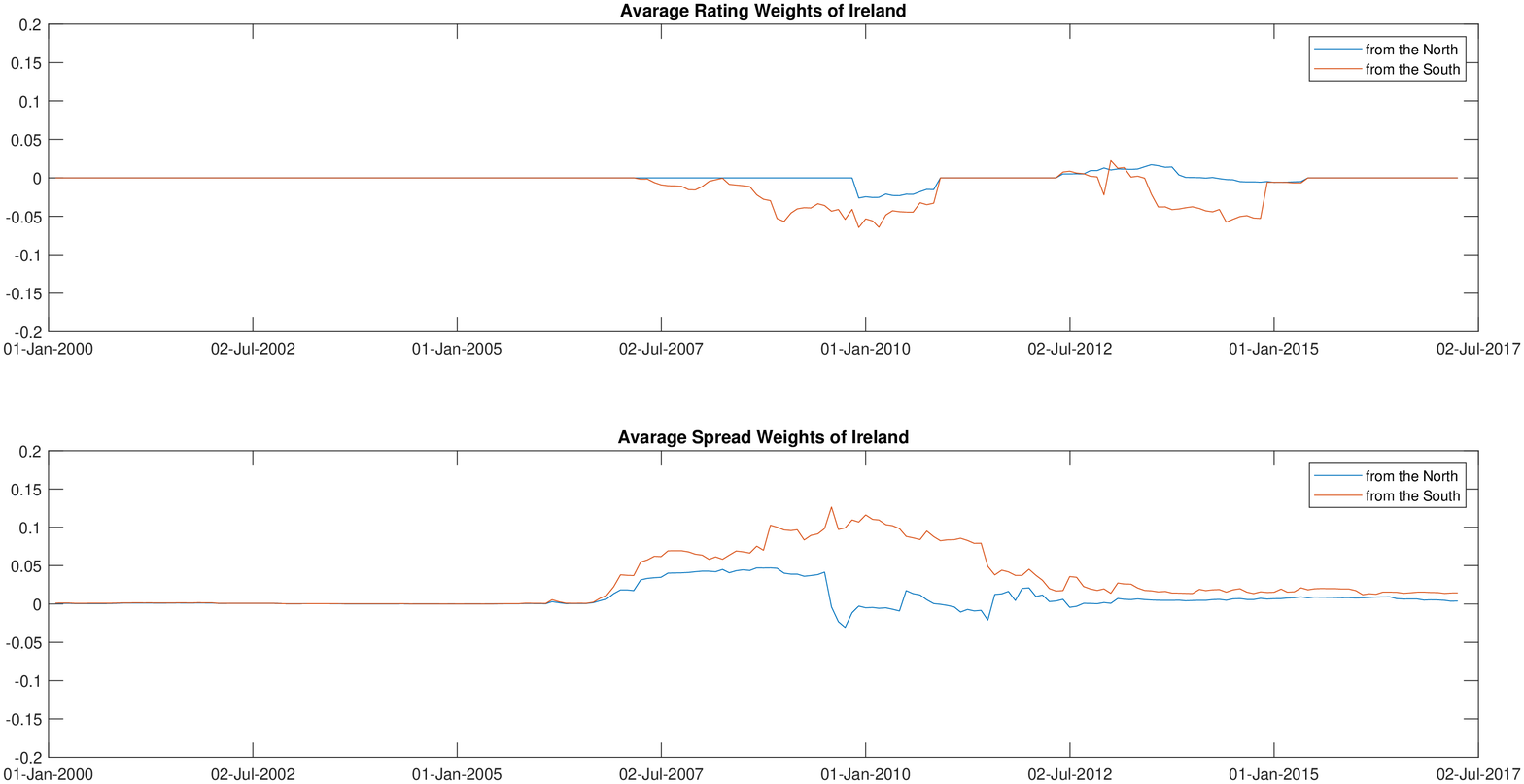}
  \\[\smallskipamount]
   \includegraphics[width=.50\textwidth]{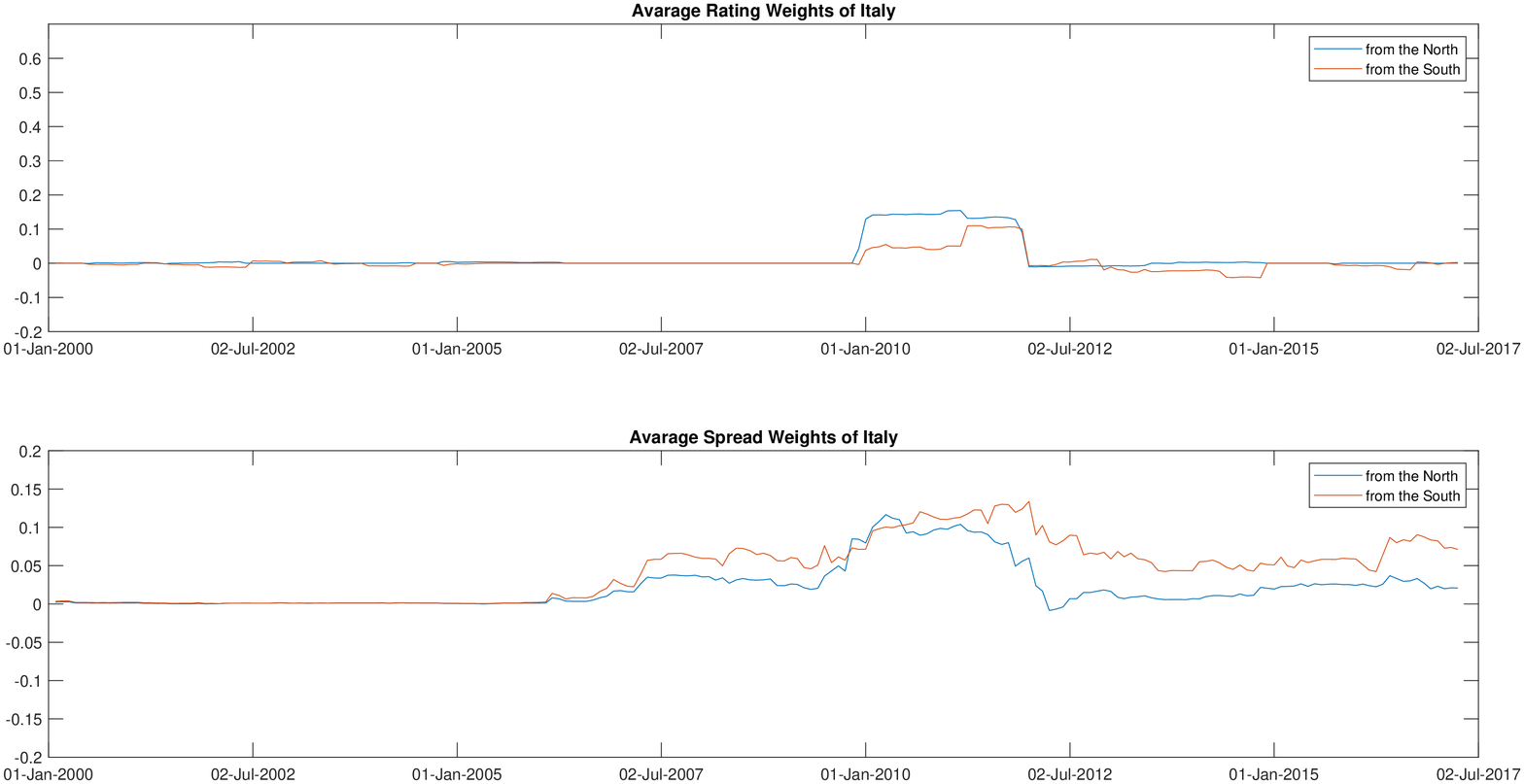}\hfill
    \includegraphics[width=.50\textwidth]{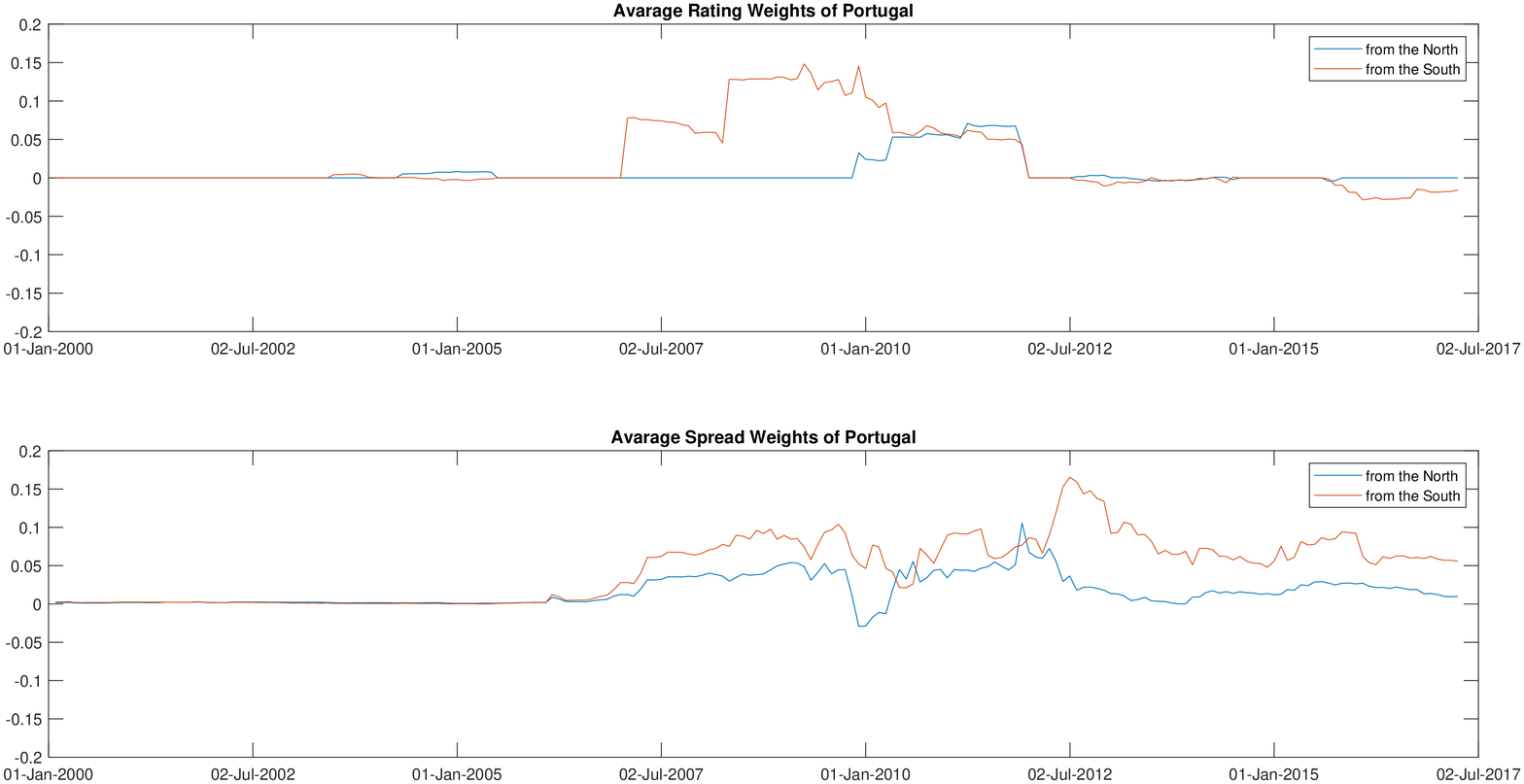}
  \\[\smallskipamount]
    \caption{Spatial structures}\label{fig:2_yr}
\end{figure}

Since model (\ref{euro_model}) is dynamic, the spatial spillovers measured by impulse responses are complicated by time dependence. To better understand the spatial structure, we
next plot the average spatial weights associated with each country in Figure 2. For the $i^{th}$ country, the average rating spatial weights from the south is computed by taking
the mean of the non-zero elements in $W_{i}^{rate}$ that are associated with the southern euro area countries; and the average rating spatial weights from the north is computed by
taking the mean of the non-zero elements in $W_{i}^{rate}$  that are associated with the northern euro area countries. In the same way, we calculate the average spread spatial
weights from the south and the north. Plots in Figure 2 provides further evidence for the marked differences in how a country is spatially linked with the north and the south. In
general, northern countries are less spatially influenced by other countries than the southern countries, warranting we taking a closer look at how a southern country's rate and
spread are spatially related to those of other countries.

In terms of the spatial relationships between one country's ratings and the ratings of other countries, Greece and Ireland stand out during the crises period, with other countries'
ratings spatially influencing these two countries' ratings negatively, highlighting the limitations of traditional spatial weights matrix where all the elements are non-negative.
Apart from Greece and Ireland, a country's sovereign ratings tend to be positively associated with other countries's ratings. Interestingly, ratings of Spain and Portugal are more
closely associated with those of other southern countries than those of the northern countries from 2007 to 2010, but become more influenced by northern countries' ratings in the
euro area crises after 2010. Italy's sovereign ratings, by contrast, are more spatially influenced by ratings of the northern countries than those of the south throughout time.

For spreads, negative weights are rare, indicating an increase in one country's spread tend to be associated with spread increases in other countries too. Apart from Spain, whose
spreads are more spatially influenced by the northern countries for a brief period during euro area crises, spreads of all southern countries are more closely spatially linked with
other southern countries than with the northern ones.

\section{Conclusions}
In applied work, if the spatial weights matrix set a priori were to be far from its true value, the empirical analysis using SAR models would be misleading. In this paper, we have
developed a two-stage VB approach to estimating panel SAR models with unknown spatial weights matrices so as to let the data speak. Our two-stage VB with D-L priors method can be
easily extended by using other popular priors such as the Lasso, Horseshoe and  spike and a slab priors. Furthermore, the success of two-stage VB shows the potential of combining
VB with more sophisticated methods, such as three-stage least squares,  full information likelihood and GMM to estimate panel SAR models, especially those involving $N\gg T$. Monte
Carlo experiments show that our two-stage VB is rather fast, and it can recover the spatial impacts well for both the long and short panels.

As an empirical example, we apply the two-stage VB to the sovereign bond ratings and sovereign spreads data of 10 eurozone countries to uncover the impacts of spatial spillovers of
the southern countries, which were more severely hit by the eurozone debt crises, and those of the northern countries. Without pre-imposing any spatial weights matrices that might
be unrealistic, we are able to shed new lights on the spillover behaviours of the south and the north. To our best knowledge, our findings are among the first in the literature to
delineate how an individual country is affected by the spillovers from other countries in the eurozone.

\pagebreak

\newpage
    \section*{Appendix: Data Sources}
    \renewcommand{\theequation}{A.\arabic{equation}}
    \renewcommand{\thesection}{A}
    \setcounter{equation}{0}

    \medskip

 \setcounter{bean}{0}
      \begin{list}
       {(\alph{bean})}{\usecounter{bean}}
  \item Spreads: monthly data from Statistical Data Warehouse, European Central Bank
  \item Ratings: monthly data from Fitch, Standard and Poor’s and Moody’s
  \item The ratio of government debt to GDP: quarterly data from Thomson Reuters Datastream
  \item Real GDP growth: quarterly data from Thomson Reuters Datastream
  \item Current account balance as a percentage of GDP: Either monthly or quarterly data from Thomson Reuters Datastream
  \item Harmonised Consumer Price Index: monthly data from Thomson Reuters Datstream
  \item Political uncertainty: quarterly data from IFO World Economic Survey. Two variables have been used in order to create the variable political uncertainty. 1) The present
      climate for foreign investors, political stability (which was discontinued), which was updated with information from the new variable, 2) Political instability, using an
      appropriate mapping for the reported values of each variable.
   \end{list}



\begin{thebibliography}{99}
\addcontentsline{toc}{section}{\refname}


  \bibitem{}
          Ahrens, A. and Bhattacharjee, A.(2015).
        \newblock Two-step lasso estimation of the spatial weights matrix.
        \newblock {\em Econometrics\/} {\em 3}, 128--55.

\bibitem{}
    Anselin, L. (1988).
      \newblock {\em Spatial Econometrics: Methods and Models\/}.
      \newblock Kluwer Academic, Dordrecht.

  \bibitem{}
        Baltagi, B.~H., S.~H. Song and Koh, W. (2003).
        \newblock Testing panel data regression models with spatial error correlation.
        \newblock {\em Journal of Econometrics\/} {\em 117}, 123--50.

 \bibitem{}
        Baltagi, B.~H., Egger, P. and Pfaffermayr, M. (2013).
        \newblock A generalized spatial panel data model with random effects.
        \newblock {\em Econometric Reviews\/} {\em 32}, 650--85.

 \bibitem{}
         Basak, G.~A., Bhattacharjee, A. and Das, S. (2018).
         \newblock Causal ordering and inference on acyclic networks.
        \newblock {\em Empirical Economics\/} {\em 55}, 213--32.

 \bibitem{}
         Bhattacharya, A., Pati, D., Pillai, N.~S. and Dunson, D.~B. (2015).
        \newblock Dirichlet–Laplace priors for optimal shrinkage.
        \newblock {\em Journal of the American Statistical Association\/} {\em 110}, 1479--90.

 \bibitem{}
        Blei, D.~M., Kucukelbir, A. and McAuliffe, J.~D.(2017).
        \newblock Variational inference: A review for statisticians.
        \newblock {\em Journal of the American Statistical Association\/} {\em 112}, 859--77.

\bibitem{}
    Case, A. (1991).
        \newblock Spatial Patterns in Household Demand.
        \newblock {\em Econometrica\/} {\em 59}, 953--65.

 \bibitem{}
         Cliff, A.~D. and Ord, J.~K. (1973).
     \newblock {\em Spatial Autocorrelation\/}.
      \newblock  London: Pion.

 \bibitem{}
      Debarsy, N., Ertur, C. and LeSage, J.P., (2012).
      \newblock Interpreting dynamic space–time panel data models.
      \newblock {\em Statistical Methodology\/} {\em9}, 158--171.

 \bibitem{}
    Fox, J. (1979).
        \newblock Simultaneous equation models and two-stage least squares.
        \newblock {\em Sociological methodology\/} {\em 10}, 130--50.


 \bibitem{}
         Gibson, H.~D., Hall, S.~G. and Tavlas, G.~S. (2017).
        \newblock Self-fulfilling dynamics: The interactions of sovereign spreads, sovereign ratings and bank ratings during the euro financial crisis.
        \newblock {\em Journal of International Money and Finance\/} {\em 73}, 371--85.

 \bibitem{}
         Gibson, H.~D., Hall, S.~G., Gefang,D., Petroulas, P. and Tavlas, G.~S. (2021).
        \newblock Cross-country spillovers of national financial markets and the effectiveness of ECB policies during the euro-area crisis.
        \newblock {\em Oxford Economic Papers\/} {\em 73}, 1454--70.

 \bibitem{}
         Gefang, D., Koop, G. and Poon, A. (2020).
        \newblock Computationally efficient inference in large Bayesian mixed frequency VARs.
        \newblock {\em Economics Letters\/} {\em 191}, 109120.

\bibitem{}
         Gefang, D., Koop, G. and Poon, A. (2022).
        \newblock Forecasting using variational Bayesian inference in large vector autoregressions with hierarchical shrinkage.
        \newblock {\em International Journal of Forecasting\/}.

 \bibitem{}
     Hall, S.~G., Gefang, D. and Tavlas, G.~S. (2023).
        \newblock A test to select between spatial weighting matrices.
        \newblock {\em Journal of Spatial Econometrics\/} {\em 4}.

 \bibitem{}
Krisztin, T. and Piribauer, P. (2023).
         \newblock A Bayesian approach for the estimation of weight matrices in spatial autoregressive models.
         \newblock {\em Spatial Economic Analysis \/} {\em 18}, 44--63.



 \bibitem{}
        Krock, M., Kleiber W. and Becker, S. (2021).
        \newblock Nonstationary modeling with sparsity for spatial data via the basis graphical lasso.
        \newblock {\em Journal of Computational and Graphical Statistics\/} {\em 30}, 375--89.

 \bibitem{}
        Lam, C. and Souza, P. C. (2019).
        \newblock Estimation and selection of spatial weight matrix in a spatial lag model.
        \newblock {\em Journal of Business and Economic Statistics\/} {\em 38}, 693--710.


 \bibitem{}
        Lee, L-F. and Yu, J. (2010).
        \newblock Estimation of spatial autoregressive panel data models with fixed effects.
        \newblock {\em Journal of Econometrics\/} {\em 154}, 165--85.

 \bibitem{}
        Li, Y., Craig, B.~A. and Bhadra, A. (2019).
        \newblock The graphical horseshoe estimator for inverse covariance matrices.
        \newblock {\em Journal of Computational and Graphical Statistics\/} {\em 28}, 747--57.

 \bibitem{}
       Liu, X. and Saraiva, P. (2019).
        \newblock GMM estimation of spatial autoregressive models in a system of simultaneous equations with heteroskedasticity.
        \newblock {\em Econometric Reviews\/} {\em 38}, 359--85.

 \bibitem{}
        Loaiza-Maya, R.,  Smith, M.~S., Nott, D.~J. and Danaher P.~J. (2021).
        \newblock Fast and accurate variational inference for models with many latent variables.
        \newblock {\em Journal of Econometrics.\/} {\em 230}, 339--62.

 \bibitem{}
       Makalic, E. and Schmidt, D.~F. (2016).
        \newblock A simple sampler for the horseshoe estimator.
        \newblock {\em IEEE Signal Processing Letters\/} {\em 23}, 179--82.

\bibitem{}
       Ormerod, J.~T. and Wand, M.~P. (2010).
        \newblock Explaining variational approximations.
        \newblock {\em The American Statistician\/} {\em 64}, 140--53.

\bibitem{}
Piribauer, P., Glocker, C. and Krisztin, T.(2023).
       \newblock Beyond Distance: The Spatial Relationships of European Regional Economic Growth.
       \newblock Available at SSRN 4391999.


\bibitem{}
      Tibshirani,  R. (1996).
        \newblock Regression shrinkage and selection via the lasso.
        \newblock {\em Journal of the Royal Statistical Society,\/} Series B {\em 58}, 267--88.

\bibitem{}
     Wang, H. (2012).
        \newblock Bayesian graphical lasso models and efficient posterior computation.
        \newblock {\em Bayesian Analysis\/} {\em  7}, 867--86.

\bibitem{}
     Yang, K. and Lee, L.~F. (2017).
        \newblock Identification and QML estimation of multivariate and simultaneous equations spatial autoregressive models.
        \newblock {\em Journal of Econometrics\/} {\em 196}, 196--214.

 \bibitem{}
     Zellner, A., and Theil, H. (1962).
        \newblock Three-Stage Least Squares: Simultaneous Estimation of Simultaneous Equations.
        \newblock {\em Econometrica\/} {\em 30}, 54--78.

    \end{thebibliography}
\end{document}